\documentclass[journal=jacsat,manuscript=article]{achemso}
\usepackage[version=3]{mhchem} % Formula subscripts using \ce{}
\usepackage{xcolor}
\usepackage{csquotes}
\author{Ashley J. Tyler}
\affiliation{Diamond Quantum Sensing Hub, Faculty of Engineering, University of Nottingham, Nottingham, NG7 2RD, United Kingdom}
\alsoaffiliation[University of Nottingham]{Optics and Photonics Research Group, University of Nottingham, Nottingham, NG7 2RD, United Kingdom}
\altaffiliation{Present Address: Cerca Magnetics Limited, Nottingham, NG7 1LD, United Kingdom}
\author{Thomas D. Bateman-Price}
\affiliation{Diamond Quantum Sensing Hub, Faculty of Engineering, University of Nottingham, Nottingham, NG7 2RD, United Kingdom}
\alsoaffiliation[University of Nottingham]{Optics and Photonics Research Group, University of Nottingham, Nottingham, NG7 2RD, United Kingdom}
\altaffiliation{Present Address: Dstl, Porton Down, Salisbury, Wiltshire, SP4 0JQ, United Kingdom}
\author{Bradley T. Flinn}
\affiliation[University of Nottingham]{School of Chemistry, University of Nottingham, Nottingham, NG7 2RD, United Kingdom}
\alsoaffiliation{Diamond Quantum Sensing Hub, Faculty of Engineering, University of Nottingham, Nottingham, NG7 2RD, United Kingdom}
\alsoaffiliation[University of Nottingham]{Advanced Materials Research Group, University of Nottingham, Nottingham, NG7 2RD, United Kingdom}
\author{William J. Cull}
\affiliation[University of Nottingham]{School of Chemistry, University of Nottingham, Nottingham, NG7 2RD, United Kingdom}
\altaffiliation{Present Addresses: SuperSTEM Laboratory, Sci-Tech Daresbury Campus, Daresbury, WA4 4AD, United Kingdom; School of Physics, Engineering and Technology, University of York, Heslington, YO10 5DD, United Kingdom}
\author{Tom J. P. Irons}
\affiliation[University of Nottingham]{School of Chemistry, University of Nottingham, Nottingham, NG7 2RD, United Kingdom}
\author{Andrei N. Khlobystov}
\affiliation[University of Nottingham]{School of Chemistry, University of Nottingham, Nottingham, NG7 2RD, United Kingdom}
\author{Melissa L. Mather}
\affiliation{Diamond Quantum Sensing Hub, Faculty of Engineering, University of Nottingham, Nottingham, NG7 2RD, United Kingdom}
\alsoaffiliation[University of Nottingham]{Advanced Materials Research Group, University of Nottingham, Nottingham, NG7 2RD, United Kingdom}
\email{Melissa.Mather@nottingham.ac.uk}
\title[An \textsf{achemso} demo]
  {Quantum Sensing for Spatial Spin Noise Mapping via Nitrogen-Vacancy Magnetic Quenching in Diamond}


\begin{document}

%%%%%%%%%%%%%%%%%%%%%%%%%%%%%%%%%%%%%%%%%%%%%%%%%%%%%%%%%%%%%%%%%%%%%%%%%%%%
\begin{tocentry}
\begin{center}
    \includegraphics[width=\linewidth]{Figures/final_TOC.png}
    \label{For Table of Contents Only.}
\end{center}
\end{tocentry}
%%%%%%%%%%%%%%%%%%%%%%%%%%%%%%%%%%%%%%%%%%%%%%%%%%%%%%%%%%%%%%%%%%%%%%%%%%%%

%%%%%%%%%%%%%%%%%%%%%%%%%%%%%%%%%%%%%%%%%%%%%%%%%%%%%%%%%%%%%%%%%%%%%
%% Abstract
%%%%%%%%%%%%%%%%%%%%%%%%%%%%%%%%%%%%%%%%%%%%%%%%%%%%%%%%%%%%%%%%%%%%%
\begin{abstract}
Detecting and mapping spin noise can reveal spatial variations associated with surface defect states, catalytic residues, free radicals and spin behaviour in spintronic materials. However, bulk measurements average over spatial heterogeneity, while scanning-probe maps require sequential rastering. Measurements based on nitrogen-vacancy (NV) centres in diamond provide an alternative approach, although established NV spin-noise protocols commonly use synchronised optical pulse sequences or resonant microwave excitation. The former requires optical timing and control hardware, while microwave delivery can introduce field inhomogeneity and thermal loading. Here, we demonstrate a microwave-free magnetic quenching method for wide-field spin-noise mapping using NV centres in diamond. The method combines continuous light-emitting diode (LED) illumination with a low-frequency, amplitude-modulated magnetic field, without requiring resonant microwave excitation or pulsed optical sequences. Magnetic quenching arises from field-induced spin-state mixing, which reduces NV photoluminescence. We investigate this photoluminescence response as a function of optical excitation power and magnetic-field modulation amplitude and interpret the observed behaviour using a theoretical model of spin-state mixing. We evaluate the method using a concentration series of aqueous gadobutrol, a paramagnetic magnetic resonance imaging contrast agent, with magnetic quenching and optically detected magnetic resonance (ODMR) measurements exhibiting consistent concentration-dependent trends. We then map variations in the spin-noise response across boron nitride and single-walled carbon nanotube samples with differing defect and residual metallic catalyst contributions. Complementary microscopy, elemental, thermal and spectroscopic characterisation supports the interpretation of differences between the nanotube samples. In particular, magnetic quenching detects a spin-noise response from boron nitride nanotubes that is silent in bulk electron paramagnetic resonance (EPR) spectroscopy. The resulting capability could support spatial studies of defect-associated spin noise in quantum technologies and advanced materials, spin behaviour in spintronic systems, and free-radical distributions in biological environments.
\end{abstract}

%%%%%%%%%%%%%%%%%%%%%%%%%%%%%%%%%%%%%%%%%%%%%%%%%%%%%%%%%%%%%%%%%%%%%
%% Intro
%%%%%%%%%%%%%%%%%%%%%%%%%%%%%%%%%%%%%%%%%%%%%%%%%%%%%%%%%%%%%%%%%%%%%
\section{Introduction}
Quantum sensing with NV centres has become established in recent years, with widespread use in applications such as detecting magnetic fields \cite{mag_OG1,Sensitive_barry,new_barry} and temperature \cite{Neumann2013,Liu2023}. Importantly, the NV spin, and its coupling to environmental spins (which generate random fluctuations in their magnetic fields, or \enquote{spin noise}), can be optically detected under ambient conditions \cite{OG_simp,Steinert2013,2013_all_optical,T1_flake,2026noise}, even down to the single spin level \cite{JP_T1,Du2024}. As such, Nitrogen vacancy (NV) centres in diamond offer a promising avenue for sensing paramagnetic materials.

Paramagnetic nanomaterials, including single-walled carbon nanotubes, endohedral fullerenes, paramagnetic metal complexes and lanthanide-based nanoparticles, exhibit remarkable physicochemical and functional properties with implications for advanced research in quantum information processing, magnetic resonance imaging, spintronics, and electrocatalysis \cite{nano13030598,ma15041535,D4NR04012K,molecules29071639,D3NA01064C}. The intricate interplay of electron spins within these materials significantly influences their functionality, emphasising the importance of precise control and characterisation of their paramagnetic nature, particularly under operational conditions. To fully harness the potential of these nanomaterials, it is imperative to develop robust characterisation methods capable of probing magnetic states beyond the bulk scale and under normal operational conditions. Traditional methods, such as electron paramagnetic resonance (EPR) spectroscopy, face limitations because they primarily yield bulk information and are mainly suited to the analysis of $S = 1/2$ systems. Scanning probe technologies, such as magnetic force microscopy (MFM), offer effective measurement techniques for spatially mapping magnetic properties \cite{MFM}. However, these methods primarily concentrate on ferromagnetic properties, necessitate specialised equipment for implementation, and can be slow due to their scanning nature. Thus, alternative strategies prioritising spatio-temporal information, spin sensitivity, and operability under relevant environmental conditions are essential---a gap NV centres are well placed to fill.

While various NV sensing protocols exist \cite{Sensitive_barry,Degen_Sense}, many widely adopted ones, including Optically Detected Magnetic Resonance (ODMR), $T_1$ and $T_2$ relaxometry, rely on the delivery of microwave frequency magnetic fields for control of the NV spin state. Achieving spatially homogeneous microwave fields, critical for spatial mapping, presents a substantial challenge \cite{Mariani2020,antenna}. Moreover, the high permittivity of microwaves in numerous sample types, especially water, presents challenges in achieving significant sensing contrast \cite{MWabsorb1, MWabsorb2}. In response, efforts to develop NV sensing protocols independent of microwaves have given rise to a microwave-free approach in $T_1$ relaxometry \cite{2013_all_optical,Lukin_T1}, which is particularly relevant for NV-based spin noise sensing. However, this approach introduces challenges in hardware synchronisation, demands for a rapidly switching high-power light source, and is susceptible to errors associated with photoionisation-induced NV charge state conversion \cite{T1_Chargecon}. Relevant to spin noise sensing and beyond, growing interest surrounds the pursuit of technically simpler NV sensing methods, particularly microwave-free protocols that capitalise on the dependence of NV photoluminescence (PL) on applied static or low frequency magnetic fields. Various approaches have emerged, including those employing large on-axis magnetic fields with external perturbations to energy level crossings \cite{wickey,wickey2, PhysRevApplied.21.044039}, photoluminescence quenching observed under a 'strong' magnetic field applied oblique to the intrinsic quantisation axis of an NV centre at an angle of approximately $55^{\circ}$ \cite{Tetienne2012}, and zero-field methods incorporating amplitude modulation of a control magnetic field to eliminate orientation dependence.

Here, we present a new perspective for microwave-free NV sensing protocols utilising amplitude-modulated magnetic fields for detecting and spatially mapping magnetic spin noise in proximity to NVs within diamond. This simplified approach, herein referred to as magnetic quenching (MQ), streamlines instrumentation and enables operation in environments where microwaves may pose risks, lead to electromagnetic interference, or in samples with low dielectric permittivity. By employing a light-emitting diode (LED) at the low, continuous-wave optical power used in this work, we reduce the risk of photoionisation-induced charge state conversion. Our demonstration showcases this easily implementable technique, investigating the optical power and magnetic field strength dependence of PL quenching and compares findings with continuous wave ODMR measurements. We demonstrate the effectiveness of our approach using gadobutrol, a paramagnetic magnetic resonance imaging contrast agent, by illustrating that ODMR and MQ show correlated changes, with consistent trends with optical power and spin noise concentration. A theoretical basis for the observed PL quenching is provided to aid interpretation of experimental observations. Furthermore, we highlight the high sensitivity to proximal spin noise of magnetic PL quenching by mapping spin noise from nanostructures with varying metallic content and surface oxidation. This work advances magnetic spin noise detection in nanomaterials and more generally provides a versatile tool for characterising paramagnetic materials. Anticipated applications span analytical sciences, materials discovery, biomedical research, and nanotechnology, offering ready access to quantum sensing technologies.
%%%%%%%%%%%%%%%%%%%%%%%%%%%%%%%%%%%%%%%%%%%%%%%%%%%%%%%%%%%%%%%%%%%%%
%% Results and Discussion
%%%%%%%%%%%%%%%%%%%%%%%%%%%%%%%%%%%%%%%%%%%%%%%%%%%%%%%%%%%%%%%%%%%%%

\section{Results and Discussion}

\begin{figure}
    \includegraphics[draft=false, width=\textwidth]{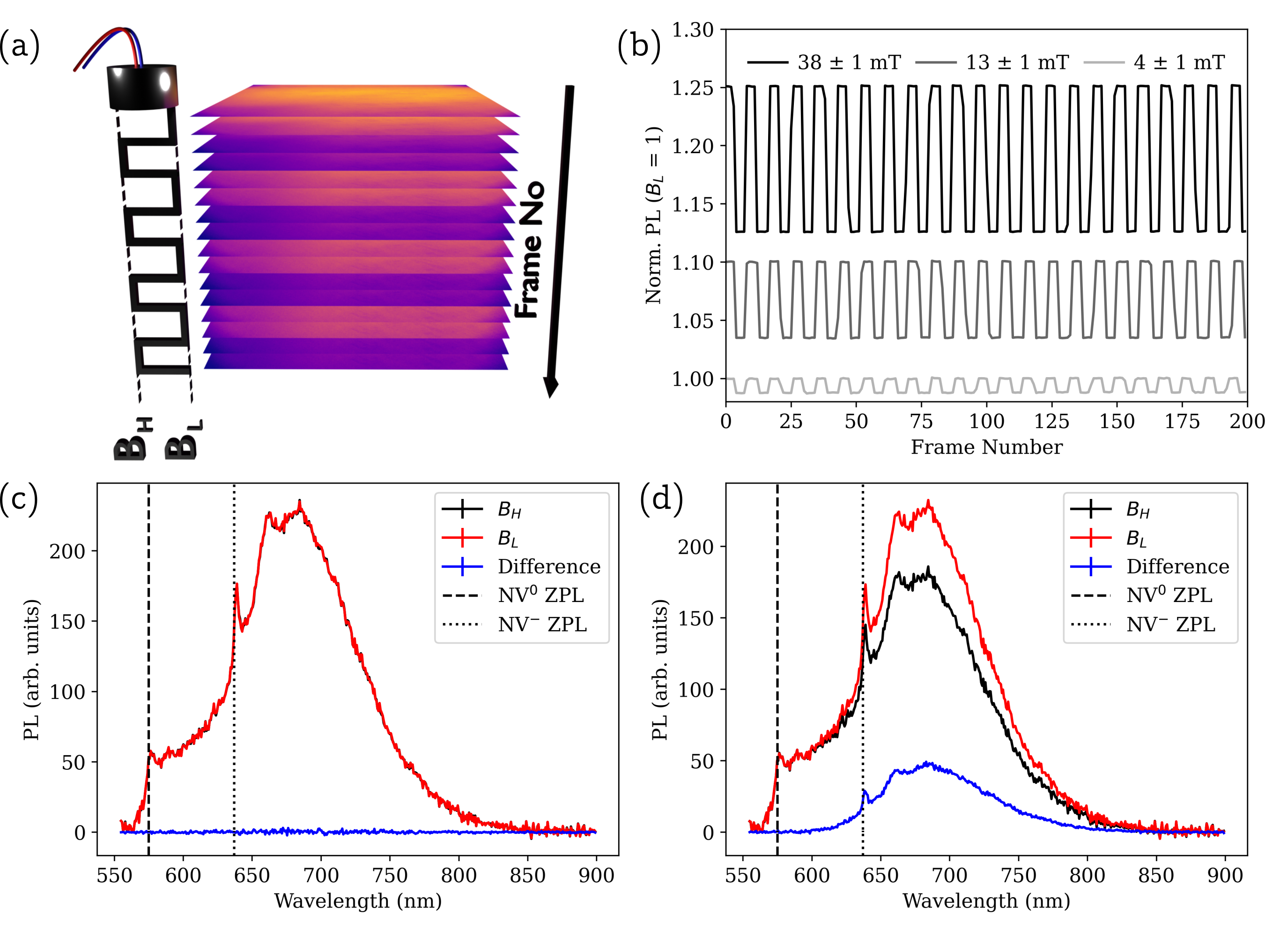}
    \caption{Magnetic quenching (MQ) imaging and spectroscopic characterisation. (a) Schematic of MQ imaging, where alternating bright and dark frames are acquired under magnet-low ($B_L$) and magnet-high ($B_H$) states. (b) Normalised PL intensity versus frame number for modulation amplitudes $(B_H-B_L)$ of 4, 13, and 38 mT, vertically offset for clarity. Intensities are normalised to the mean $B_L$ intensity. (c) PL spectra for $\leq$1 mT modulation: $B_L$ (red), $B_H$ (black), and difference $B_L-B_H$ (blue), showing minimal spectral variation. (d) PL spectra for 38 mT modulation, showing reduced PL in the $B_H$ state and a corresponding difference spectrum. These schematics and measurements introduce wide-field MQ imaging as a selective probe of NV$^{-}$ centre magnetic quenching, capable of producing $\approx$12\% difference in PL intensity between the magnet-low and magnet-high states.}
    \label{fig3}
\end{figure}

The effect of magnetic modulation amplitude on NV centre PL was investigated through MQ measurements performed on a diamond chip under ambient conditions ('bare chip'). The results are summarised in Figure \ref{fig3}. Figure \ref{fig3}(a) provides a schematic representation of the measurement protocol, depicting the sequence of camera image acquisition and applied magnetic field modulation. (b) presents normalised PL intensity as a function of frame number for magnetic modulation amplitudes of 4 mT, 13 mT, and 38 mT; each measurement is normalised to its own mean intensity in the magnet-low state ($B_L = 1$). We observed a reduction in PL intensity upon application of a magnetic field, equivalent to a maximum reduction of 12.1 $\pm$ 0.1 \% at 38 mT (6.35 $\pm$ 0.1 \% at 13 mT, 1.18 $\pm$ 0.1 \% at 4 mT). These results align with reductions in PL intensity (10\% to 20\%) reported in experiments using NV ensembles and significantly higher magnetic fields (of the order of 100 mT) combined with precise magnet-ensemble orientation control \cite{epstein,quench_fiber}. Notably, our simplified approach, employing a commercial wide-field microscope, an LED, and an off-the-shelf electromagnet, yielded comparable PL modulation capabilities to these more sophisticated experimental techniques.

To confirm that the observed PL changes in our MQ sensing protocol are primarily due to NV$^{-}$ centres - the negative charge state of NV centre, commonly utilised for quantum sensing applications - spectrophotometric measurements of the emitted light were conducted. The resulting PL emission spectra, obtained from the bare chip under magnet-low ($B_L$) and magnet-high ($B_H$) states, along with the difference ($B_L-B_H$) spectrum, are presented in Figure \ref{fig3}(c) and (d). Figure \ref{fig3}(c) displays spectra acquired at a low magnetic modulation amplitude ($\leq$ 1 mT). Both magnet-low and magnet-high state spectra exhibit zero phonon lines (ZPL) characteristic of NV$^{0}$ (the neutral charge state) at 575 nm and NV$^{-}$ at 637 nm along with corresponding phonon sidebands. The difference spectrum in Figure \ref{fig3}(c) shows negligible change across the measured wavelength range. In contrast, Figure \ref{fig3}(d) shows spectra acquired at a higher magnetic modulation amplitude (38 mT). Again, the magnet-low and magnet-high states reveal characteristic ZPLs and phonon sidebands. However, a reduction in PL intensity is evident in the magnet-high state spectrum, which is more clearly observed in the difference spectrum. Notably, the difference spectrum shows a significant change specifically within the spectral region attributed to NV$^{-}$ emission, with minimal change at the NV$^{0}$ ZPL. 
Quantitatively, the change in the 550 nm to 637 nm band in the difference spectrum of Figure \ref{fig3}(d), which is dominated by NV$^{0}$ emission, accounts for only 4.7 $\pm$0.1 \% of the total emission. This observation, combined with the fact that the spectral changes are small and predominantly localised near the NV$^{-}$ ZPL, demonstrates that our MQ technique leverages changes in the NV$^{-}$ PL as the primary readout mechanism.

\begin{figure}
    \includegraphics[width=\linewidth]{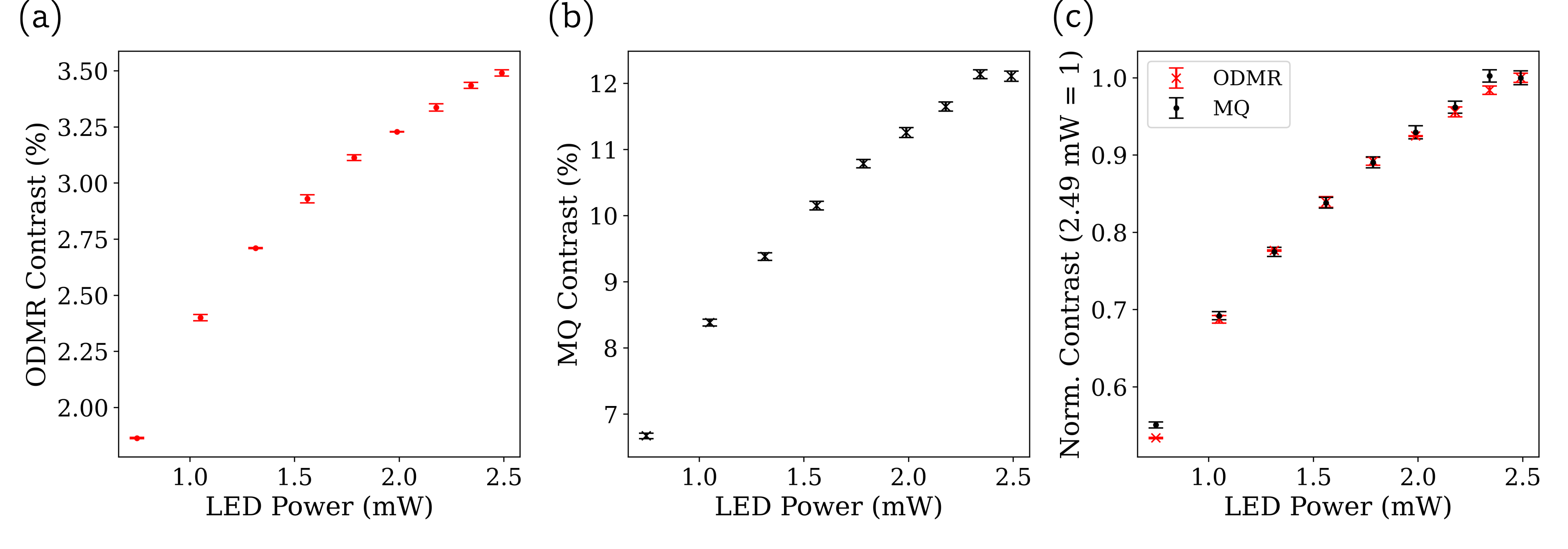}
    \caption{Optical excitation power dependence of ODMR and MQ contrast. ODMR (a) and MQ contrast (b) as a function of optical excitation power, measured at the sample plane. (c) ODMR and MQ contrast normalised to their respective contrast value at 2.49 mW, the maximum optical excitation power used in this setup. ODMR and MQ contrast show a similar dependence on optical excitation power.}
    \label{fig4}
\end{figure}

Figure \ref{fig4} examines the relationship between ODMR and MQ contrast as a function of optical excitation power, building upon previous research on the power dependence of ODMR contrast \cite{dru}. Throughout this work, contrast is defined as the percentage change in PL intensity induced by the application of resonant microwaves (ODMR) or application of the external magnetic field (MQ), relative to the PL intensity in the corresponding off state - off-resonance microwaves for ODMR, and the magnet-low state for MQ. We measured both ODMR and MQ contrast at ten distinct optical excitation powers, ranging from 0.75 mW to 2.49 mW. Figure \ref{fig4}(a) displays the ODMR contrast, determined from a double Lorentzian fit, and Figure \ref{fig4}(b) shows the MQ contrast (38 mT modulation amplitude), across the same excitation power range. To facilitate direct comparison, both contrasts were normalised to one at the maximum excitation power of 2.49 mW, as shown in Figure \ref{fig4}(c).

The observed positive correlation between optical excitation power and contrast in both ODMR and MQ indicates that optical spin polarisation within the NV centre ensemble remains unsaturated within the investigated power range, consistent with the use of our relatively low-power LED source. This can be understood by considering the structure of the NV triplet ground state $^{3}A_{2}$ with its three spin sub-levels, $\vert 0 \rangle$ and $\vert \pm 1 \rangle$. Under ambient conditions, there are near-equal equilibrium populations within each state, as dictated by the Boltzmann distribution, and a zero-field splitting of $D_{gs}=2.87$ GHz. Continuous optical excitation leads to an increased population in the $\vert 0 \rangle$ sub-level and decreased populations in the $\vert \pm 1 \rangle$ sub-levels, owing to the higher rate of spin non-conserving, non-radiative decay from the $\vert \pm 1 \rangle$ components of the triplet excited state $^{3}E$ relative to the $\vert 0 \rangle$ component and the spin conserving nature of optical transitions. Because optical excitation and emission are probabilistic processes, a single cycle does not result in complete ground state spin polarisation. Instead, the final spin polarisation state is determined by the cumulative effect of multiple excitation and emission events, which depend on the optical source's power. Thus, the observed increase in contrast with optical power indicates that saturation was not reached for either ODMR or MQ. Indeed, PL intensity is maximised at saturation spin polarisation due to the minimised non-radiative decay rate from the $\vert 0 \rangle$ component of the $^{3}E$ state. The contrast in ODMR and MQ arises from the change in spin polarisation induced by microwave resonance (ODMR) or magnetic field application (MQ). Since the population decay rate from the initial state depends exponentially on the initial population, the contrast depends on LED power within the range investigated here. This process is common to both ODMR and MQ, explaining the closely aligned trends in contrast with LED power shown in Figure \ref{fig4}(c).

Considering the absolute values of contrast obtained for ODMR and MQ (Figures \ref{fig4}(a) and (b)), it is seen that MQ contrast is significantly higher ($\sim$12 \%) than the ODMR contrast ($\sim$3.5 \%), despite the similar trend with increasing LED intensity. This difference is primarily attributed to the application of an off-axis external magnetic field in the MQ method. In our setup, the applied magnetic field is perpendicular to the $(100)$ plane of the diamond chip and results in an angle of approximately $55^{\circ}$ between the field and the NV centre quantisation axis. This off-axis field induces spin state mixing, where the components of the triplet states $\vert\tilde{i}\rangle$ are no longer pure spin states but become linear combinations of the zero-field spin sub-levels, as discussed in the Methods section. This spin state mixing reduces the difference in non-radiative decay rates between excited triplet state components, diminishing the significance of the higher photoluminescence intensity resulting from ground-state triplet component excitation \cite{Welter2022}. In other words, this spin state mixing alters the population distribution across many spin sublevels which enhances the difference in PL between the magnet-low and magnet-high states. Conversely, ODMR relies on resonant microwave transitions between specific spin sub-levels. The efficiency of these transitions, and thus the achievable contrast, is inherently limited by factors such as field inhomogeneity from the microwave antenna, reducing the selectivity and efficiency of driving the desired transitions, microwave absorption within the sample, which further weakens the microwave field strength and, for continuous wave ODMR, effect of optical spin repolarisation.

\begin{figure}
    \includegraphics[width=\linewidth]{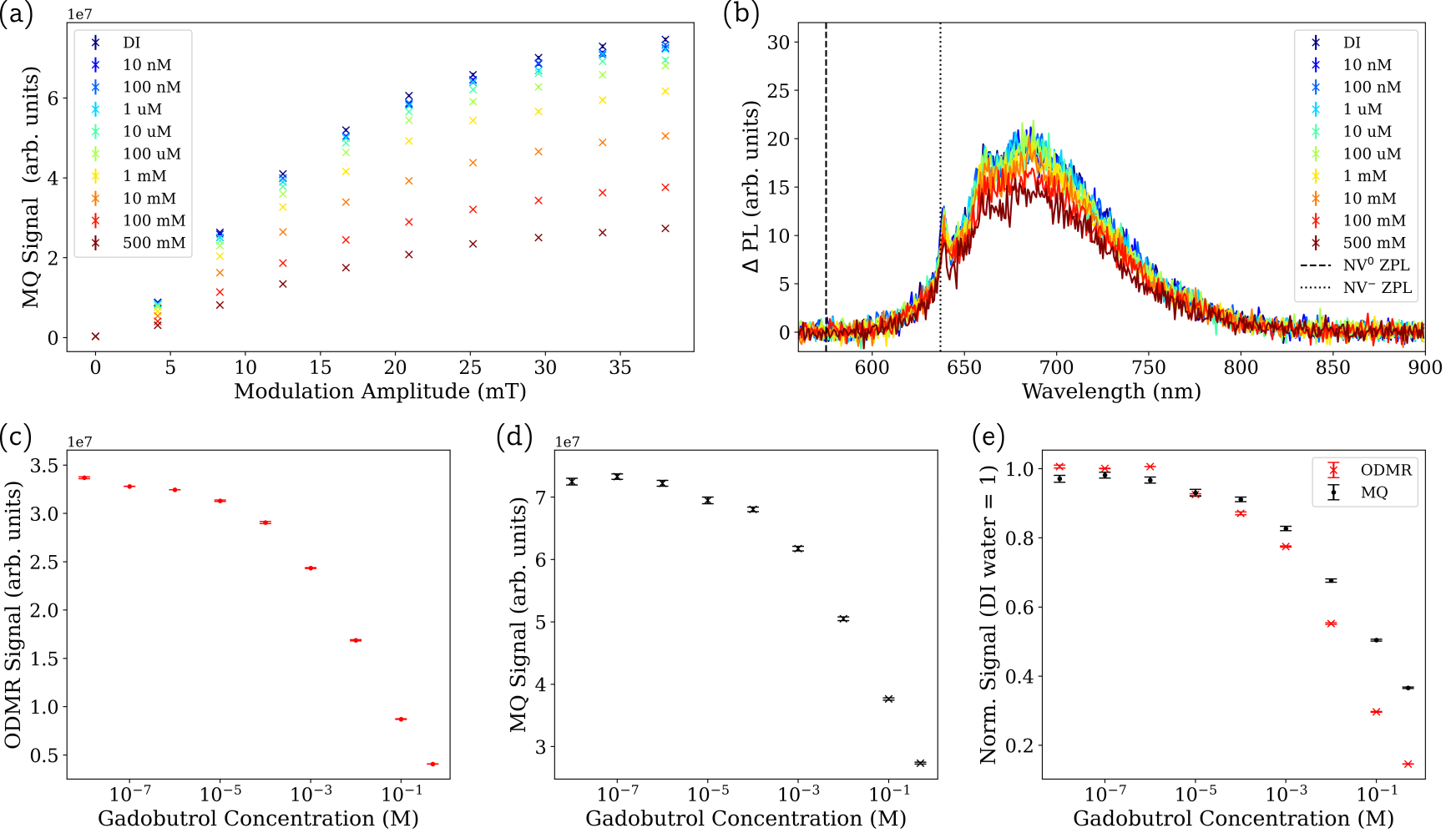}
    \caption{MQ and ODMR sensing of controlled spin-noise concentrations. Measurements of a diamond chip following the addition of solutions with increasing gadobutrol concentrations. (a) MQ signal as a function of gadobutrol concentration and magnetic modulation amplitude. (b) Spectral profile of the MQ signal, confirming that the measured response is dominated by emission NV$^{-}$ centres around the 637 nm zero-phonon line. (c) ODMR signal and (d) MQ signal as a function of gadobutrol concentration. (e) ODMR and MQ signal at each concentration, normalised to the respective signal recorded for the deionised water control. Both signals decrease with increasing gadobutrol concentration, demonstrating that MQ reproduces the concentration-dependent response of ODMR for spin-noise sensing.}
    \label{fig5}
\end{figure}

To demonstrate the MQ sensing protocol's ability to detect spin noise from paramagnetic materials, we performed proof-of-principle measurements. Specifically, we introduced controlled spin noise at the diamond surface using aqueous solutions of gadobutrol, a paramagnetic Magnetic Resonance Imaging (MRI) contrast agent containing Gd$^{3+}$ ions, at concentrations ranging from 0 mM (deionised water) to 500 mM. This allowed us to observe the impact of increasing environmental spin noise on the MQ signal. The MQ signal, derived from the difference in average integrated PL intensity between magnet-low state and magnet-high state camera frames, shows a positive correlation with the magnetic modulation amplitude (Figure \ref{fig5}(a)). This indicates that stronger magnetic fields enhance the MQ response, due to increased spin state mixing. The differences in MQ signal across varying gadobutrol concentrations become more pronounced at these higher modulation amplitudes, suggesting SNR could be increased further with an increased modulation amplitude.

To gain further insight into the spectral characteristics of the MQ signal and its response to gadobutrol, we performed spectrophotometry. The difference spectra ($B_L-B_H$) reveals a prominent ZPL at 637 nm, characteristic of NV$^{-}$ emission, with minimal contribution from NV$^0$ (Figure \ref{fig5}(b)). This confirms that the observed MQ signal originates primarily from NV$^{-}$ centres, consistent with their known sensitivity to magnetic fluctuations. The dominance of NV$^{-}$ emission in the difference spectrum indicates that our technique selectively probes the spin dynamics of NV$^{-}$ centres, enabling interpretation of signal changes as a function of gadobutrol concentration as arising from the fluctuating magnetic fields generated by Gd$^{3+}$ ions.

To directly compare the induced change of ODMR and MQ by the paramagnetic spin noise from gadobutrol, we performed sequential measurements. The ODMR signal, representing the change in integrated PL intensity upon resonant microwave application, and the MQ signal, representing the change in PL intensity upon application of an external magnetic field, were plotted as a function of gadobutrol concentration (Figure \ref{fig5}(c) and \ref{fig5}(d)). To facilitate direct comparison, the ODMR and MQ signals were each normalised to their respective signal recorded using the deionised water control, shown in Figure \ref{fig5}(e). Both techniques show a generally similar trend of decreasing signal with increasing gadobutrol concentration, further supporting the idea that both methods are sensitive to paramagnetic spin noise. However, subtle differences in their responses provide valuable mechanistic information.

The interaction between Gd$^{3+}$ ions and NV$^{-}$ centres, which underlies this sensitivity to paramagnetic spin noise, is primarily governed by the magnetic dipolar interaction. The Gd$^{3+}$ ions, with their seven unpaired electrons (spin $S=\frac{7}{2}$), generate a fluctuating magnetic field. This interaction, described in Equations S1-S7 of the Supplementary Information (SI), depends on the distance between the Gd$^{3+}$ ions and the NV centre. Increasing the concentration of the solution acts to decrease the average distance between Gd$^{3+}$ at the diamond-solution interface and an NV centre located at depth, $d$, below the diamond surface. This results in a non-linear increase in strength of the fluctuating magnetic field with concentration (Equation S5 and Figure S1c). In addition, the power spectral density of the fluctuating magnetic field is broadened with increasing concentration (Equation S6, Figure S1b). The non-linear decrease in ODMR and MQ signal, with increasing concentration, arises due to the interplay between the change of field magnitude and spectral profile of the fluctuation magnetic field, which both increase the NV$^{-}$ centre's longitudinal relaxation rate ($\Gamma_1$, Equation S7), and correspondingly decrease its $T_1$ time.

While the decrease in signal with increasing gadobutrol concentration observed under both ODMR and MQ is attributed to an decrease in the longitudinal relaxation time $T_1$ of the NV$^{-}$ centres,\cite{Steinert2013,2013_all_optical,gad2,Lukin_T1,gorrini_gad} as detailed in the SI, the underlying mechanisms by which these techniques detect this increased relaxation, and thus generate a measurable signal from the difference between two acquisition states - magnet-high/magnet-low for MQ and off-/on-resonance microwaves for ODMR - differ \cite{brad1,Brad2,Brad3,Brad4,sensorspaper}. Specifically, MQ relies on magnetic field-induced spin state mixing, while ODMR utilises microwave-driven transitions between spin sublevels.

In MQ, the application of a magnetic field oblique relative to the NV centre's quantisation axis at an angle of approximately $55^{\circ}$ results in spin state mixing (illustrated in the Theoretical Methods section - Figure \ref{fig:energy_levels}). This mixing reduces the difference in non-radiative decay rates between excited triplet state components, thus diminishing the achievable spin polarisation and the MQ signal. Furthermore, as discussed in the Theoretical Methods section, the applied magnetic field in the MQ method breaks the degeneracy of the $\vert \pm 1 \rangle$ components of the ground state. This lifting of degeneracy leads to spin noise-induced relaxation from the lowest energy level ($\vert0 \rangle$) primarily into the $\vert+1 \rangle$ level, which can influence the rate of relaxation.

In contrast, ODMR relies on resonant microwave transitions between specific spin sublevels. In the absence of an external field, the $\vert \pm 1 \rangle$ components of the ground state remain degenerate. This degeneracy allows relaxation from the lowest energy component to occur into both $\vert \pm 1 \rangle$ levels, potentially increasing the overall spin relaxation rate. The equations governing these transitions and relaxation rates are detailed in the Supplementary Information. These mechanistic differences likely contribute to the variations observed in the concentration-dependent signal changes between MQ and ODMR (Figure \ref{fig5}(e)). As predicted by the theoretical treatment in the Supporting Information and established in previous NV $T_1$ relaxometry studies of gadolinium-containing systems \cite{JP_T1,Steinert2013,gad2}, increasing gadobutrol concentration increases the magnitude and broadens the spectral density of the magnetic noise generated by Gd$^{3+}$ ions. Both effects increase the NV-centre longitudinal relaxation rate, $\Gamma_1 = 1/T_1$, reducing the achievable spin polarisation and consequently decreasing the MQ and ODMR signals. Although direct $T_1$ relaxometry was not performed, the concentration-dependent trends observed using both modalities are consistent with this relaxation-mediated contribution to the sensing response. By contrast, the underlying MQ contrast is generated by magnetic-field-induced spin-state mixing and the associated modification of radiative and non-radiative decay pathways \cite{Tetienne2012,Choi2012,Goldman2015}. MQ therefore constitutes a distinct optical spin-noise sensing modality rather than a direct implementation of $T_1$ relaxometry. Direct $T_1$ measurements would provide an additional independent test of the relaxation contribution; nevertheless, the agreement between the theoretical treatment and the concentration-dependent MQ and ODMR responses provides convergent support for the proposed interpretation of the spin-noise sensing results.

In summary, the gadobutrol measurements demonstrate that MQ responds systematically to controlled changes in paramagnetic spin noise, with concentration-dependent trends corresponding to those observed using ODMR. The two methods employ distinct readout mechanisms, with MQ based on magnetic-field-induced spin-state mixing and ODMR on resonant microwave-driven transitions, while both responses are influenced by spin-noise-induced longitudinal relaxation.

\begin{figure}
    \includegraphics[width=\linewidth]{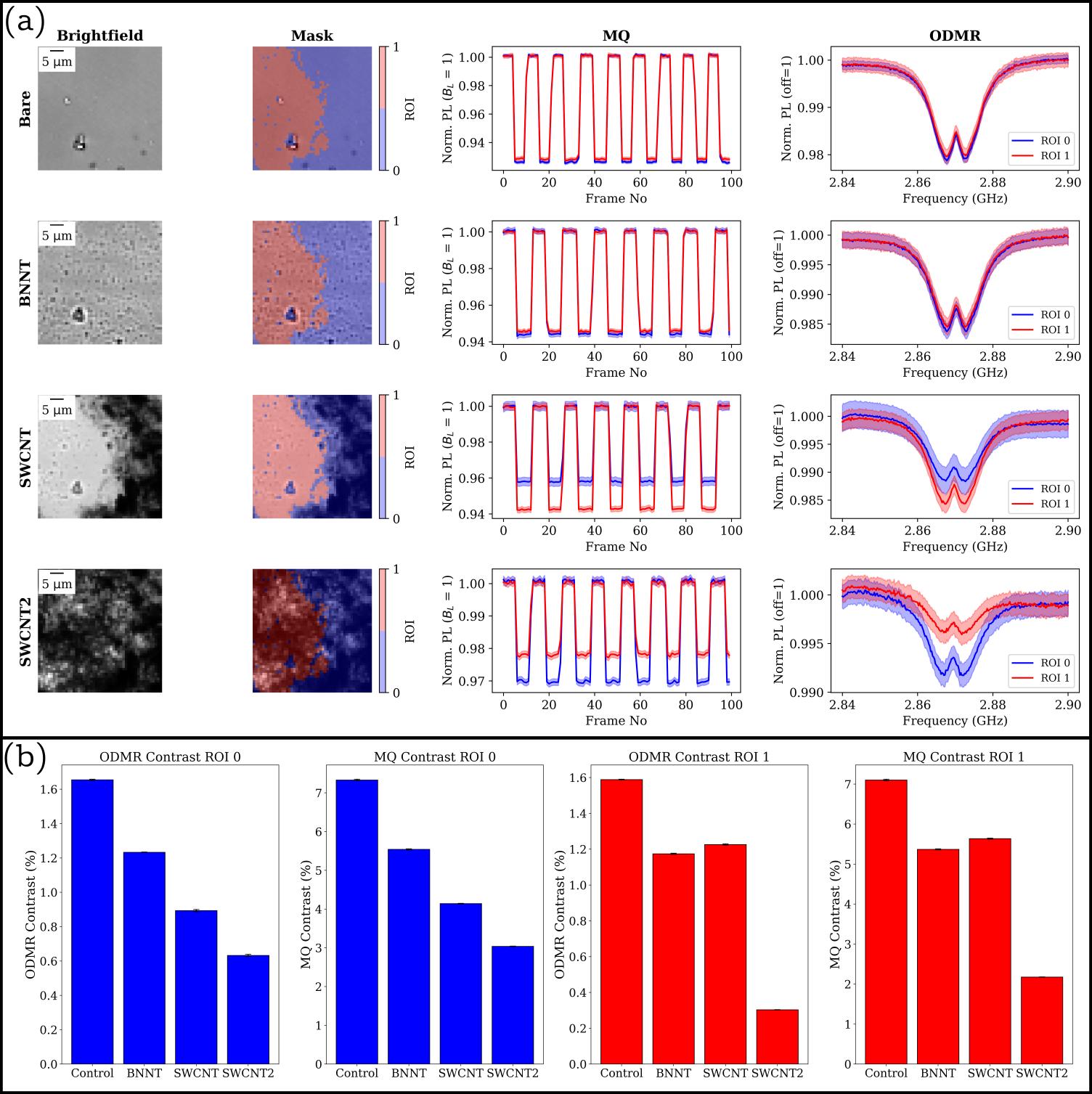}
    \caption{Spatially resolved MQ and ODMR sensing of nanotube spin noise. Correlated ODMR and MQ measurements of the same diamond region following sequential deposition of BNNTs, SWCNT1, and SWCNT2. (a) Brightfield images (column 1), ROI segmentation masks (column 2), and MQ (column 3) and ODMR (column 4) measurements for the bare diamond control and after each deposition. ROI masks were defined from the SWCNT1 deposition, where the nanotubes formed a spatially distinct pattern, and the same masks were applied to ROI 0 and ROI 1 across all samples. MQ data are normalised to the average magnet-low-state PL intensity per ROI; ODMR data to the off-resonance PL intensity. (b) ODMR and MQ contrast for ROI 0 and ROI 1 following each deposition. Both techniques show consistent spatially dependent changes in contrast across the deposition sequence, with SWCNT2 producing the largest reduction, demonstrating the spatial sensitivity of MQ to local spin noise.}
    \label{half_spatial}
\end{figure}

\begin{table}[bt]
  \centering
  \caption{ODMR and MQ signals for each ROI (normalised to the signal of the bare diamond control)}
  \label{tbl:odmr_mq}
  \begin{tabular}{lcc|cc}
    \hline
    \textbf{Nanotube} & \multicolumn{2}{c|}{\textbf{ROI 0}} & \multicolumn{2}{c}{\textbf{ROI 1}} \\
                   & ODMR & MQ & ODMR & MQ \\
    \hline
    BNNT    & 0.75$\pm$0.01 & 0.76$\pm$0.01 & 0.74$\pm$0.01 & 0.76$\pm$0.01 \\
    SWCNT1  & 0.54$\pm$0.01 & 0.56$\pm$0.01 & 0.77$\pm$0.01 & 0.79$\pm$0.01 \\
    SWCNT2  & 0.38$\pm$0.01 & 0.41$\pm$0.01 & 0.19$\pm$0.01 & 0.31$\pm$0.01 \\
    \hline
  \end{tabular}
\end{table}

Previous studies using nanodiamond sensors have demonstrated the detection of spin noise from solid-state materials \cite{brad1,Brad2}. Here, we extend MQ sensing to a planar diamond platform to investigate spatial variations in the response to weakly paramagnetic nanomaterials deposited at the diamond surface. We investigated single-walled carbon nanotubes (SWCNTs) and boron nitride nanotubes (BNNTs), in which material composition and processing influence spin-active defect and residual catalyst contributions. These nanomaterials are used in nanoelectronics, composites, and drug delivery, where local spin environments can influence material function. For chemical vapour deposition (CVD)-grown carbon nanotubes, the type and amount of residual metal left over from synthesis can have a significant effect on the observed level of paramagnetism, while BNNTs typically host fewer paramagnetic sites owing to their metal-catalyst-free synthesis.  To support the interpretation of the MQ and ODMR sensing results presented herein, additional chemical analyses were performed on all nanotube species (see Figures S2, S3 and S4). Electron paramagnetic resonance (EPR) spectroscopy, high-resolution transmission electron microscopy (HRTEM), energy dispersive X-ray analysis (EDX), thermogravimetric analysis (TGA) and powder X-ray diffraction (PXRD) were utilised to characterise the different nanotube compositions and identify sources of spin noise. Both varieties of SWCNT were shown to contain residual metal catalyst from their synthesis. The SWCNT with lower metal content (5.1 \%, Ni) is herein designated as SWCNT1, and the SWCNT with higher metal content (16.4 \%, Fe) as SWCNT2.  

Figure \ref{half_spatial}(a) presents sequential ODMR and MQ measurements on a diamond chip: bare (control), boron nitride nanotubes (BNNT), SWCNT1 and finally SWCNT2. Samples were added in increasing order of expected spin noise contribution. The columns depict: brightfield images, Region of Interest (ROI) masks (two regions, each approximately 1000 $\mu$m$^{2}$), normalised MQ, and normalised ODMR measurements. For the control (bare chip) and BNNT samples, ODMR and MQ contrast were similar between ROI 0 and ROI 1, reflecting sample uniformity. However, BNNT deposition resulted in a measurable reduction of both contrasts by approximately 25\% when compared to the bare diamond control measurement. As detailed in the Methods, we report this reduction as signal normalised to the bare diamond control signal rather than as comparing contrast levels (see Table \ref{tbl:odmr_mq}, which shows ODMR and MQ signal as a fraction of the bare diamond control signal for both ROIs), to preserve common-mode noise rejection across the sequential depositions. The reduction in MQ and ODMR signals is consistent with a change in the local spin environment following BNNT deposition, despite the weak paramagnetic character of the BNNT sample. The source of this paramagnetic response is potentially defects or surface states, as suggested by Panich \textit{et al.}\cite{Panich2005} EPR spectroscopy confirmed on a bulk scale BNNT's near-negligible spin concentration compared to SWCNTs (Figures S2, S3 and S4). Sonication of BNNTs in IPA, employed to ensure uniform sample distribution, may have inadvertently introduced mechanical defects into the BNNTs, which could subsequently incorporate carbon impurities, as previously observed in hexagonal boron nitride (hBN) \cite{Li2023}.

SWCNT1 deposition significantly reduced ODMR and MQ contrast in ROI 0, while contrast from ROI 1 remained less affected, consistent with a greater nanotube-associated spin-noise contribution in ROI 0 due to SWCNT1. EPR spectroscopic analysis revealed a broad peak, consistent with observations by Cambré \textit{et al.} \cite{Cambre2009}, (see also Figure S3), attributed to a combination of residual metal catalyst, conduction electrons and defect-trapped electrons, all contributing to the observed spin noise. ROI 1's slight contrast change compared to BNNT results suggests potential partial removal of BNNT by the solvent the SWCNT1 sample was dispersed in, accompanied with limited SWCNT1 deposition. SWCNT2, with higher metal loading (16.4 \%, Fe compared to 5.1 \% Ni - see Figures S3 and S4), reversed the contrast pattern. ROI 1 now exhibited lower contrasts than ROI 0. We hypothesise that residual SWCNT1 in ROI 0 acted as a barrier, spatially offsetting SWCNT2 from the NV centres. Conversely, in ROI 1, SWCNT2 directly covered the surface. The difference in NV response between the ROIs demonstrates spatial mapping of nanotube-associated variations in the MQ response. The correspondence with ODMR and the independently characterised differences in residual metal content and bulk paramagnetic character support interpretation of these variations as arising from differences in the local spin-noise environment.

\begin{figure}
    \includegraphics[width=\linewidth]{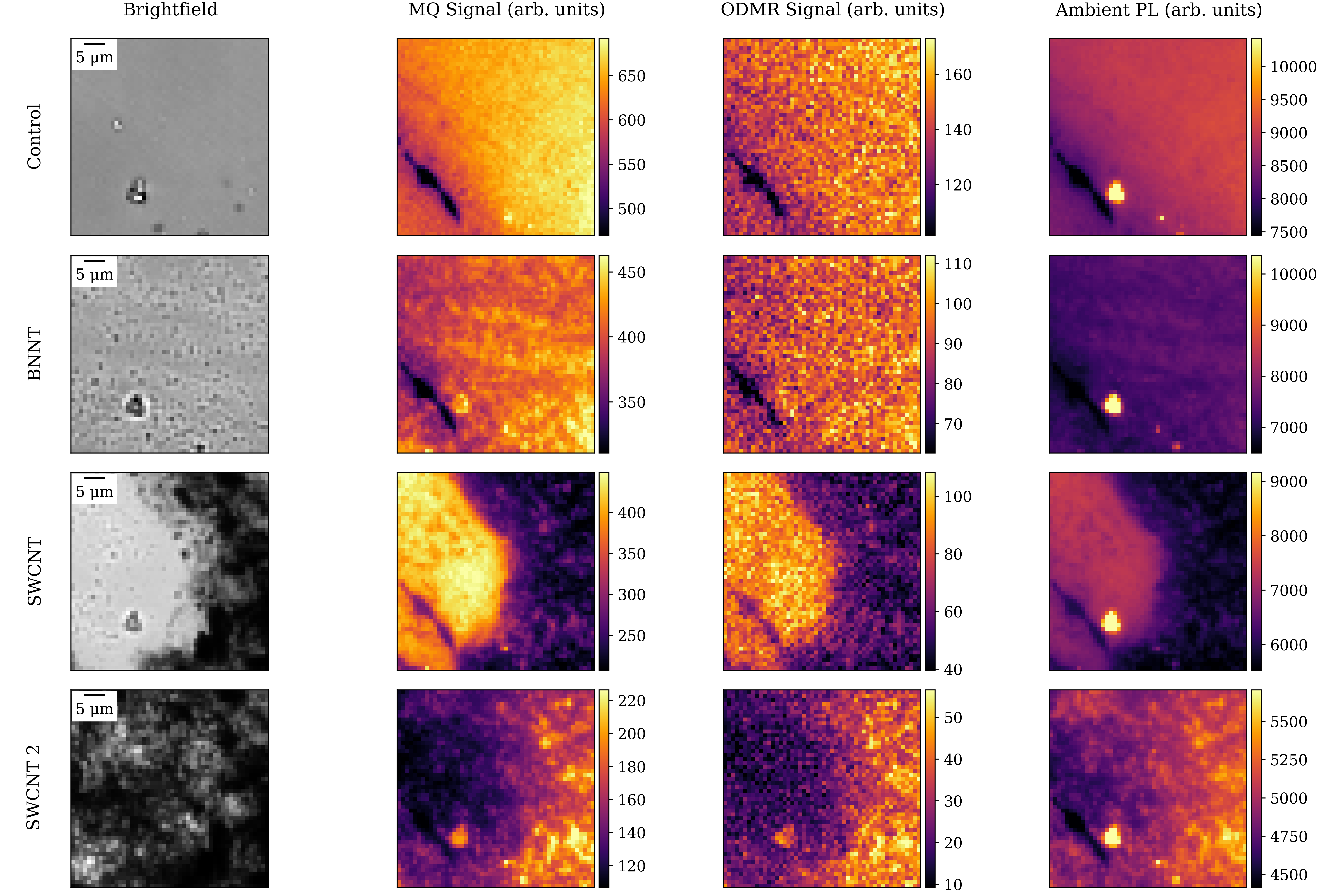}
    \caption{Spatial mapping of local spin noise using MQ and ODMR. Brightfield images (column 1), MQ maps (column 2, raw MQ signal) and ODMR maps (column 3, raw ODMR signal), and ambient PL images (column 4), acquired from the same region of the diamond sensor following sequential deposition of BNNTs, SWCNT1 and SWCNT2. The MQ and ODMR maps exhibit similar spatial patterns, revealing nanotube-dependent variations in local spin noise that are not readily apparent in the corresponding brightfield or ambient PL images. The strongest spatial variation is observed following deposition of SWCNT1 and SWCNT2, demonstrating the ability of both sensing modalities to spatially resolve local variations in the spin-noise environment.}
    \label{fig6}
\end{figure}

To transition from ROI-based analysis to finer spatial detail, we employed pixel-wise MQ and ODMR signal mapping (Figure \ref{fig6}, columns 2 \& 3), with colourmaps restricted to the 0.5$^\text{th}$ and 99.5$^\text{th}$ percentiles. Alongside, we included brightfield images and PL images obtained without magnetic or microwave fields (Figure \ref{fig6}, columns 1 \& 4), showcasing light collected using a 575 nm longpass filter. The brightfield and PL images show a consistent contaminant feature. However, this feature is significantly reduced in MQ and ODMR signal maps, demonstrating effective decoupling of background light from NV$^{-}$ PL  achievable using both methods. Despite this suppression, the feature appears in a subset of MQ and ODMR signal maps as a region of relatively high signal, specifically in MQ (BNNT and SWCNT2) and ODMR (SWCNT2). Notably, this feature is absent in the corresponding signal maps from the bare diamond control, suggesting it is magnetically silent, generating no spin noise. Instead, we hypothesise that it acts as a physical barrier, increasing the distance between the NV centres beneath the feature and the deposited nanotubes. This increased separation reduces the interaction strength, thereby highlighting MQ sensing’s enhanced sensitivity to surface-correlated spin noise, particularly in the presence of high spin densities. For SWCNT2, our MQ and ODMR signal maps revealed localised variations in PL quenching that were imperceptible in brightfield images, further confirming the presented methodologies ability to resolve subtle, spatially-dependent spin noise interactions directly at the diamond surface. Overall, the distinct spatial patterns observed across PL, MQ, and ODMR maps allow us to differentiate regions with subtle variations in paramagnetic properties. Furthermore, the superior signal-to-noise ratio attained using MQ, evident in the reduced spatial variation of colour compared to ODMR, reinforces the value of the simple measurement protocol MQ offers.

\section{Conclusion}
We have demonstrated an NV-centre sensing protocol based on magnetic quenching (MQ), in which a low-frequency, amplitude-modulated magnetic field produces a measurable change in photoluminescence. Measurements as a function of optical excitation power and magnetic-field modulation amplitude, supported by spectroscopic characterisation and theoretical analysis, support a mechanism in which field-induced spin-state mixing modifies the radiative and non-radiative decay pathways of the NV centre. Differential measurement between magnet-low and magnet-high states also provides common-mode rejection of background photoluminescence variations. Comparative measurements using MQ and optically detected magnetic resonance (ODMR) showed corresponding trends with optical excitation power and controlled changes in spin noise introduced using aqueous gadobutrol solutions. The concentration-dependent responses are consistent with an increase in the NV longitudinal relaxation rate caused by the increased magnitude and broadened spectral density of the magnetic noise generated by Gd$^{3+}$ ions. These results show how changes in the surrounding spin-noise environment modulate the MQ response, while its underlying optical contrast arises from field-induced spin-state mixing. We further applied MQ to map spatial variations in the spin-noise response across boron nitride nanotube and single-walled carbon nanotube deposits. MQ and ODMR exhibited corresponding spatial patterns and differentiated samples with differing defect and residual metallic catalyst contributions. Complementary EPR spectroscopy, transmission electron microscopy, energy-dispersive X-ray spectroscopy, thermogravimetric analysis and powder X-ray diffraction provided structural, compositional and magnetic information that supported interpretation of the differences between the nanotube samples. In particular, MQ detected a spin-noise response following deposition of boron nitride nanotubes despite their near-silent bulk EPR spectrum.Together, these results establish MQ as a distinct method for detecting and mapping spatial variations in spin-noise responses across heterogeneous samples. While direct $T_1$ measurements could provide an additional quantitative test of the relaxation-mediated response, the convergence of our theoretical treatment, ODMR benchmarking, and corresponding concentration-dependent trends supports the underlying sensing mechanism. Future work can build upon this framework to establish quantitative limits of detection and performance benchmarks across diverse operating conditions. Ultimately, this approach provides a framework for spatial investigations of defect-associated spin noise in quantum technologies and advanced materials, spin behaviour in spintronic systems, and free-radical distributions in biological environments.

%%%%%%%%%%%%%%%%%%%%%%%%%%%%%%%%%%%%%%%%%%%%%%%%%%%%%%%%%%%%%%%%%%%%%
%% Methods
%%%%%%%%%%%%%%%%%%%%%%%%%%%%%%%%%%%%%%%%%%%%%%%%%%%%%%%%%%%%%%%%%%%%%

\section{Methods}\label{sec:methods}

\subsection{Experimental Methods}
\subsubsection{Materials}

Two subsets of NV-diamond chips were used as sensing platforms, referred to as \textbf{Type 1} and \textbf{Type 2}. Both chip types were electronic-grade, single-crystal diamond with a (100) crystal orientation and nitrogen concentration below 5~ppb, grown by chemical vapour deposition (Element Six, UK).

\textbf{Type 1} chips were used for gadobutrol titrations (Figure~\ref{fig5}) and spatial imaging experiments (Figures~\ref{half_spatial} and~\ref{fig6}). These chips were implanted with \textsuperscript{14}N\textsuperscript{+} ions at 4~keV and a dose of $10^{13}$~ions/cm\textsuperscript{2} (Innovion Corp., USA), followed by high-temperature annealing and surface oxidation using methods described by Chen \textit{et al.}~\cite{Chen2017}.

\textbf{Type 2} chips were used for the optical power measurements presented in Figures~\ref{fig3} and~\ref{fig4}. They were implanted with an energy of 6~keV and a dose of $10^{13}$~ions/cm\textsuperscript{2} (\textsuperscript{14}N isotope), annealed at high temperature and surface oxidised (Qnami, Switzerland).

Prior to experimentation, all diamond chips underwent a cleaning procedure to remove surface contaminants; this process involved sequential acid treatments, neutralisation, rinsing, and annealing (detailed in the SI). Following cleaning, the diamond chips were fixed to 25 mm x 25 mm x 0.17 mm glass microscope coverslips using an optical adhesive (Norland Optical Adhesive 61) and cured under UV light. The resulting diamond chip-coverslip assemblies were then secured to a custom-designed printed circuit board (PCB) with cellulose nitrate lacquer to enable microwave delivery for ODMR measurements.

Ten aqueous solutions of gadobutrol were prepared by adding a known concentration of gadobutrol (Gadobutrol monohydrate, 160 mg, MilliporeSigma\textsuperscript{\textregistered}, Burlington, MA, USA; SKU:Y0001803) to deionised water. The concentrations ranged from 0 mM to 500 mM, and 200 $\mu$L of each solution was applied to the surface of the diamond plate for measurements. Between the application of each gadobutrol solution, the diamond chip was washed with deionised water. The solutions were applied in order of increasing concentration.

Solid state samples included boron nitride nanotubes (BNNT) and single-walled carbon nanotubes with different levels and types of residual metal catalyst. Specifically, one type with low metal loading (SWCNT1 - 5.1 \% Ni), and a second type with higher metal loading (SWCNT2 - 16.4 \% Fe). Nanotube properties were characterised using Transmission Electron Microscopy (TEM), Energy Dispersive X-ray spectroscopy (EDX), Thermogravimetric Analysis (TGA), Powder X-ray Diffraction (PXRD), and Electron Paramagnetic Resonance (EPR) spectroscopy (see SI for details). BNNTs were acquired from BNNT LLC (USA). For application to the diamond surface, BNNT solutions consisting of 1 mg/mL of BNNT diluted in isopropyl alcohol (IPA) were prepared. SWCNT1s, acquired from Carbon Solutions Inc. (P2 SWCNT; USA), and SWCNT2s, acquired from Nanointergris (HiPCO SWCNT; Canada), were dispersed in isopropyl alcohol (IPA) at a concentration of 1 mg/mL for application to the diamond surface. All solutions were sonicated prior to drop casting 5 $\mu$L onto the diamond surface. MQ and ODMR sensing experiments were performed 5 minutes after drop casting which provided sufficient time for the solvent to evaporate leaving nanotube deposited on the diamond surface with no residual solvent.

\subsubsection{Instrumentation}

The experimental setup is designed to perform both ODMR and MQ protocols using an inverted fluorescence microscope (Olympus IX83) fitted with a standard microscope filter cube (Olympus IX3-FGWXL) for control of excitation and emission wavelength ranges. For illumination, a CoolLED-pE4000 system generates green excitation light (see Figure S5c for spectral profile). This light is filtered via the excitation filter (passband 530 nm to 550 nm) and reflected towards the sample by a 570 nm dichroic mirror before being focused into the back focal plane of objective lens (Olympus APO N, x60, NA = 1.49). The photoluminescence (PL) emitted from the NVs is collected by the same objective lens and passes back through the dichroic mirror, and then through a 575 nm long-pass emission filter. 

For camera-based detection of PL, an sCMOS camera (Photometrics, USA, Prime 95B) connected to the microscope's camera port via a c-mount adapter was used. The microscope's field of view is sensor-limited, resulting in a maximum 220 $\mu$m x 220 $\mu$m field of view. For spectral analysis, the PL is redirected towards the microscope's eyepiece port, and a compact spectrometer (Ocean Optics, USA, QEPRO-XR) is used. In this configuration, a 10x objective lens (Mitutoyu, Japan) relays the light to the spectrometer through a 100 $\mu$m core linear keyed fiber (Ocean Optics, USA, PL100-2-VIS-NIR).

The illumination intensity at the sample was controlled by adjusting the LED drive current, with the optical power measured using an Excelitas X-cite power measurement system (Excelitas Technologies, USA).  To convert measured optical power to optical power density, the measured power was divided by the area over which uniform sample illumination was observed. Here, we chose the sensor 220 $\mu$m by 220 $\mu$m sensor limited field of view as the area over which the measured power was distributed (0.0484 mm$^{2}$), consistent with our experimental observations. All measurements were performed at the maximum LED current corresponding to a measured power of 2.49 mW  (power density of 5.14 W/cm$^{2}$), unless otherwise stated. We note that in our inverted microscope utilising K\"ohler illumination, with an objective lens with x60 magnification and field number of 22, selecting our sensor limited field of view as the illuminated area is likely an underestimate of the area and therefore an overestimate of the power density. All optical power values stated in Figure \ref{fig4} correspond to the measured total power, without division by a factor of 0.0484 mm$^{2}$ applied for conversion to power density.

For MQ measurements, an electromagnet (holding magnet, 20 V, outer diameter = 29.5 mm, inner core diameter = 11 mm, length = 30 mm) positioned above the sample generates an amplitude-modulated magnetic field. The current supplied to the electromagnet is controlled using an LED current driver (Thorlabs, USA, T-Cube$^{\text{TM}}$) and an arbitrary function generator (Tektronix, USA, AFG 3022B) to achieve this modulation. The strength and profile of the generated magnetic field depend on the relative positioning of the NV sensing layer and the electromagnet, with this positioning controlled by a linear translation stage (Thorlabs, USA, MTS25/M-Z8). The maximum magnetic flux density that could be generated by the electromagnet in the imaging position was measured as 38 mT using a Hirst GM08 handheld Gaussmeter equipped with a TP002 transverse Hall Probe. See SI for conversion of modulation voltage to magnetic flux density.   

ODMR measurements are performed using a copper wire (thickness 0.125 mm) to deliver microwave radiation in the 2.84 GHz to 2.90 GHz range. These microwaves are synthesised using a Keysight N5182B (Keysight Technologies, USA) and amplified by a high gain amplifier (Microwave Amplifiers Ltd., UK, AM4-2-43-43), resulting in an input power of 10 W supplied to the input terminal of the PCB. The entire setup is controlled by a computer using a combination of custom Python scripts and open-source software packages (PycroManager, $\mu$manager, PyVisa). In particular, the PycroManager package provided programmatic access to $\mu$manager's hardware abstracted scientific instrument interfacing and data handling functionality \cite{pycro}. Where instrument control was not available through PycroManager/$\mu$manager, such as for interfacing with the Tektronix AFG and Keysight N5128B, the PyVisa package was used for hardware control.

Solid state samples were prepared for transmission electron microscopy by sonication in propan-2-ol and subsequent drop-casting onto lacey-carbon-coated copper TEM grids. TEM imaging was conducted using a JEOL 2100F FEG-TEM microscope operated at 200 kV at the Nanoscale and Microscale Research Centre (nmRC), University of Nottingham. 

Local EDX spectra were acquired for samples mounted on lacey-carbon-coated copper TEM grids using an Oxford Instruments INCA X-ray microanalysis system. 

A PANalytical X’Pert Pro diffractometer was used for the powder X-ray measurements. This was achieved using a Cu K$\alpha$ radiation source ($\lambda$ = 1.5432 Å, 40 kV, 40 mA) in a Bragg–Brentano geometry on a Si zero-background holder. Nanotube samples were drop-cast onto the holder with propan-2-ol. The parameters for a typical experiment were the following: 0.0525° step size, 0.00220°/s scan speed, 5° start angle, 100° stop angle, and 6080 s time/step. 

A TA Q500 Thermogravimetric Analyzer was used for the thermogravimetric analysis. All samples were analysed using a platinum pan and in the presence of air. Experimental parameters were as follows: 10 min isothermal hold at room temperature, ramp from room temperature to 1000 °C at 10 °C/min, followed by a final 10 min isothermal hold at 1000 °C. 

EPR spectra were recorded on a Bruker EMX spectrometer using Quartz glass tubes at room temperature in the X-band.

\subsubsection{ODMR and MQ Protocols}
ODMR measurements are performed by sweeping the frequency of the microwave radiation across the NV centre's zero-field splitting (approximately 2.87 GHz). The microwave frequency, generated by a Keysight N5182B, is swept in the 2.84 GHz to 2.90 GHz range, in 0.5 MHz increments. During these measurements, the PL is recorded at each frequency point using the sCMOS camera with a 20 ms exposure time. The camera controlled timing, triggering the microwave generator to the next frequency on the falling edge of a TTL signal after each frame's exposure.

MQ measurements involve modulating the magnetic field generated by the electromagnet. The current supplied to the electromagnet is square-wave modulated at 1 Hz (using a Tektronix AFG 3022B function generator) to generate a corresponding modulation of the magnetic field. The resulting magnetic field amplitude is varied up to approximately 38 mT. PL is measured using the sCMOS camera with a 20 ms exposure time. In these measurements, camera frames are classified as \enquote{magnet-high state} or \enquote{magnet-low state}. For spectrophotometry measurements, a 350 ms integration time spectrum was acquired each time the electromagnet changed state. The second channel of the signal generator, which controlled the magnetic modulation, also generated the trigger for the spectrometer. The timing diagram for the MQ spectrophotometry is shown in the SI.

For solution state measurements, the diamond chip underwent a specific sequence. Immediately after adding the solution, three ODMR sweeps were performed. The electromagnet was approached towards the sample programmatically, to an 'imaging position', 3 mm above the surface of the diamond plate. MQ measurements were performed with the full range of magnetic modulation amplitudes outlined in the MQ sensing section above. Following camera-based measurements, the process was repeated on a cleaned chip for spectral measurements.

\subsubsection{Data Analysis}

For ODMR data from studies of the bare chip and solution state samples, that lack spatial features, all the pixel values in a given frame are summed to obtain a single intensity value per frame. The ODMR spectra presented in the figures correspond to the average of 3 individual spectrum. The ODMR signal is extracted by fitting a double Lorentzian function to each individual spectrum using the Levenberg-Marquardt method accessed using the LMFIT python package. The ODMR signal stated is then the average of the signal of the two Lorentzian terms as determined by curve fitting.

For MQ sensing experiments, frames with 20 ms exposure time were continuously acquired. Each frame was later classified as either being \enquote{magnet-high} or \enquote{magnet-low}. For the bare chip and solution state samples, all the pixel values in a given frame are summed to give a single value per frame. A frame was classified as a \enquote{magnet-low} frame if the pixel value for that frame was greater than the mean of all frame values for each measurement; frames with a measured intensity lower than the mean were classified as \enquote{magnet-high}. Following frame classification, the mean average of all on (and off frames) was determined and the extent of PL quenching was calculated by subtracting the mean on value from the mean off value. Uncertainties in the measurements are expressed as the propagated standard error of the mean, calculated by numerical quadrature from the uncertainties in the mean off and on frames. Frame classification was performed on a pixel-wise basis for spatial mapping.

Where ODMR and MQ contrast are expressed as a percentage, each measurement is normalised to the ambient PL intensity for each individual measurement. For ODMR, the normalisation constant is the average off-resonance PL intensity, and for MQ, it is the average intensity as measured during the magnet's low state. This provides an intuitive measure of magnetic/RF-induced reduction in measured PL intensity. Throughout this work, we also report ODMR and MQ as the raw differential signal - the difference in PL intensity between the two measurement states (on-resonance/off-resonance for ODMR, magnet-high/magnet-low for MQ) in arbitrary units, without normalisation - here, robustness to common-mode noise rejection is prioritised over the intuitive percentage measure. Where measurements are compared across differing conditions, the raw differential signal is instead normalised to a reference sample stated in parentheses at first use (e.g. "DI water = 1", "bare diamond control = 1"). Unlike contrast, which normalises each measurement to its own internal on/off baseline, this preserves common-mode noise rejection between differing experimental conditions.

We report whichever of these representations is most appropriate for the context. Figure \ref{fig4} presents contrast, since the discussion there concerns the absolute magnitude of the change between the two measurement states in the context of the underlying mechanistic response. Figure \ref{half_spatial} similarly presents contrast, which has the added benefit of allowing both ROIs to be plotted on the same axis and their patterns compared directly, despite differing absolute PL levels between regions. Elsewhere, we report either the raw differential signal or the raw signal normalised to a stated experimental condition, as described above.

\subsection{Theoretical Methods}

\begin{figure}[h!]
    \centering
    \includegraphics[width=\textwidth]{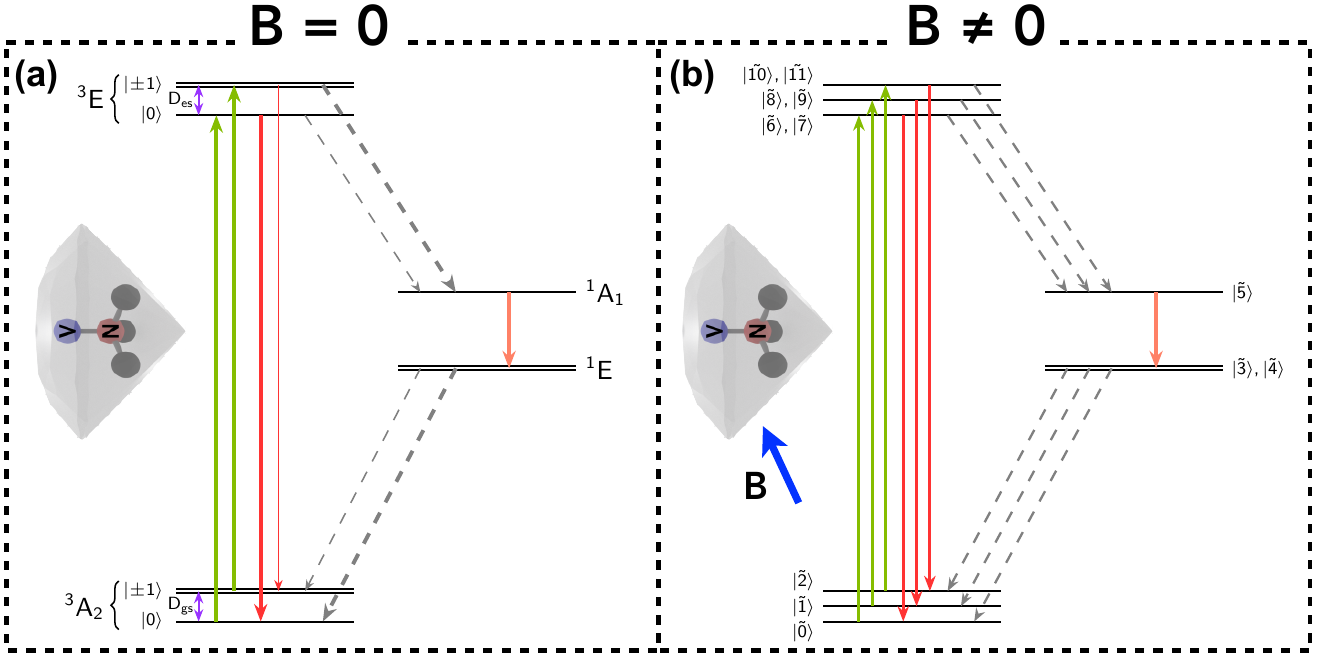}
    \caption{Electronic structure and magnetic-field-induced spin mixing of the NV$^{-}$ centre. (a) under ambient conditions and (b) a magnetic field oblique to its intrinsic quantisation axis at an angle of approximately $55^{\circ}$.~\cite{Gali2019} The ground state and third excited states, characterised as ${}^{3}A_{2}$ and ${}^{3}E$ at zero field respectively, have spin multiplicity of 3, whilst the first and second excited states, E and $A_{1}$ at zero field respectively, have spin multiplicity of 1. Optical absorptions between triplet states are given by solid green lines and radiative decay by solid red lines, whilst non-radiative decay between triplet and singlet states is shown by dashed lines and infrared radiative decay between singlet states by a brown line. At zero field (a), the three components of ${}^{3}A_{2}$ and ${}^{3}E$ are characterised by their magnetic quantum numbers $M_{s}$, denoting the spin projection along the NV centre quantisation axis, with the $M_{s}=0$ and $M_{s}=\pm1$ components of ${}^{3}A_{2}$ and ${}^{3}E$ separated by zero-field splittings of $D_{gs}$ and $D_{es}$ respectively. Under a magnetic field oblique to the NV centre quantisation axis (b) however, spin is no longer quantised along this axis and the three components of each triplet state are represented by a linear combination of zero-field states, resulting in spin-state mixing. This mixing modifies the relative rates of radiative and non-radiative decay pathways, reducing the efficiency of optical spin polarisation into the ground-state $M_{s}=0$ sublevel. Consequently, the application of an off-axis magnetic field decreases the photoluminescence intensity of the NV$^{-}$ centre, forming the basis of the magnetic quenching (MQ) sensing mechanism.}
    \label{fig:energy_levels}
\end{figure}

The operating principle of our proposed simple NV sensing protocol is fundamentally based upon detecting changes in NV PL intensity between two experimentally controlled conditions. In the first (low magnetic state), the NV centres are subject to ambient conditions. In the second state (high magnetic state) target NVs are subject to a relatively large magnetic field on the order of milliteslas applied oblique to the NV centre's intrinsic quantisation axis at an angle of approximately $55^{\circ}$. The electronic energy levels of the NV$^{-}$ centre are shown at zero field in Figure \ref{fig:energy_levels}(a) and under an external magnetic field oblique to the NV$^{-}$ centre intrinsic quantisation axis in Figure \ref{fig:energy_levels}(b).

At zero field, the NV$^{-}$ centre has $C_{3v}$ symmetry with the spin triplet $(S=1)$ ground state ${}^{3}A_{2}$. The first and second excited states, the doubly-degenerate ${}^{1}E$ and the ${}^{1}A_{1}$ states respectively, are both spin singlets $(S=0)$, whilst the third excited state is the doubly-degenerate spin triplet ${}^{3}E$ \cite{Razinkovas2021}.  At zero field, the three components of each triplet electronic state are characterised by their magnetic quantum numbers $M_{s}$ and denoted by $|SM_{s}\rangle$, where $M_{s}$ is their spin projections along the NV centre axis of spin quantisation, $\hat{S_{z}}|SM_{s}\rangle=M_{s}|SM_{s}\rangle$. The degeneracy of these spin sublevels is lifted due to the effects of spin-orbit coupling and, most significantly in the NV$^{-}$ centre, spin-spin dipolar coupling. This zero-field splitting (ZFS) of spin sublevels, along with the effect of an external magnetic field $\mathbf{B}$, can be represented by a phenomenological spin Hamiltonian \cite{Sinnecker2006,Segawa2023,Mostafanejad2014}

\begin{equation}
\hat{\mathcal{H}}=\hat{\mathbf{S}}\cdot \mathbf{D} \cdot\hat{\mathbf{S}}+\mu_{B}\mathbf{B}\cdot \mathbf{g}\cdot\hat{\mathbf{S}}
\end{equation}

where $\mathbf{D}$ is the ZFS tensor, $\hat{\mathbf{S}}$ is the spin operator, $\mu_{B}$ is the Bohr magneton and $\mathbf{g}$ the gyromagnetic tensor.

In the absence of a magnetic field, the spin Hamiltonian for the NV$^{-}$ centre reduces to $\hat{\mathcal{H}}=D[\hat{S}_{z}^{2}-\frac{1}{3}S(S+1)]$, where $D$ is the axial ZFS parameter derived from $\mathbf{D}$ (described in the SI); it follows that the Hamiltonian matrix $\langle SM_{s}|\mathcal{H}|SM_{s}^{\prime}\rangle$ is diagonal and the spin states $|SM_{s}\rangle$ are eigenstates of the spin Hamiltonian. In the presence of a magnetic field not aligned to the axis of quantisation, the Hamiltonian in the basis of the spin sublevels is no longer diagonal and its eigenstates become linear combinations of the zero-field spin states, $|\tilde{i}\rangle$, as shown in Figure \ref{fig:energy_levels}(b) and given in Equation S9.

The transitions between electronic energy levels at zero field are shown in Figure \ref{fig:energy_levels}(a); optical excitations from ${}^{3}A_{2}$ to ${}^{3}E$, broadband photoluminescence upon de-excitations from ${}^{3}E$ back to ${}^{3}A_{2}$, along with non-radiative transitions from ${}^{3}E$ to ${}^{1}A_{1}$ and ${}^{1}E$ to ${}^{3}A_{2}$. The optical excitation and radiative decay processes are spin conserving whilst the non-radiative decay processes between singlet and triplet states are spin non-conserving; these inter-system crossing (ISC) transitions, enabled by spin-orbit coupling and electron-phonon interactions, are highly state selective and occur from ${}^{3}E^{\pm1}\rightarrow{}^{1}A_{1}$ at a far higher rate than from ${}^{3}E^{0}\rightarrow{}^{1}A_{1}$. As such, the population accumulation into the lowest-energy component of the ground state is much less efficient under a magnetic field, with a higher rate of non-radiative decay from the lowest-energy components of the excited triplet state relative to zero field, thus decreasing the photoluminescence intensity \cite{Tetienne2012}.

This theoretical framework underscores how MQ sensing leverages the magnetic field-dependent photoluminescence of NV centres to probe local spin environments. Fundamentally, the MQ technique relies on comparing PL intensity between a high NV spin polarisation state (under low magnetic field) and a state of significant spin state mixing (under a high magnetic field), with changes reflecting environmental interactions. A comprehensive derivation of the spin Hamiltonian, detailed mathematical treatment of spin mixing, and a more exhaustive description of the energy levels and transition rates are provided in the SI. 

\begin{acknowledgement}
M.L.M., A.J.T. and T.D.B-P wish to acknowledge the European Research Council (ERC) for funding through the ERC Consolidator Award, TransPhorm (Grant No. 683108). M.L.M. and A.N.K. wish to acknowledge funding from EPSRC through the New Horizons scheme (Grant No. EP/V049623/1). M.L.M., B.T.F. and T.J.P.I. acknowledge funding from the Royal Academy of Engineering through their Chair in Emerging Technologies Scheme (Grant No. CiET-2223-102). B.T.F. acknowledges the support of the EPSRC doctoral prize programme. B.T.F. and W.J.C. acknowledge the Nanoscale and Microscale Research Centre (nmRC) for access to instrumentation JEOL 2100+ TEM, under Grant No. EP/L022494/1, OneView and K3-IS Gatan cameras were used under Grant No. EP/WOO6413/1. A.N.K. acknowledges funding of the EPSRC Program Grant “Metal Atoms on Surfaces \& Interfaces (MASI) for Sustainable Future” (Grant No. EP/V000055/1). A.N.K., M.L.M. and W.J.C. acknowledge funding from the Leverhulme Trust (Grant No. RPG-2022-300: “Taming the Radicals: Highly Reactive Species Incarcerated in Carbon Cages”). T.J.P.I. and M.L.M. acknowledge funding from the Nottingham Quantum Initiative Pump-Priming Funding scheme, University of Nottingham (Theoretical insights into diamond NV centre quantum magnetometry) and Nottingham Impact Accelerator: EPSRC IAA Funding Opportunities, University of Nottingham (Unlocking The Potential Of Solid-State Spin Systems With Advanced Theoretical Modelling). The authors would like to acknowledge Dr E. Stephen Davis for his assistance with performing EPR analysis. The authors would like to thank BNNT LLC for access to BNNT samples. 
\end{acknowledgement}

\begin{suppinfo}

\end{suppinfo}

\bibliography{MQ_Refs}

\end{document}

\renewcommand{\thefigure}{S\arabic{figure}}
\renewcommand{\thetable}{S\arabic{table}}
\renewcommand{\theequation}{S\arabic{equation}}
\setcounter{figure}{0}
\setcounter{table}{0}
\setcounter{equation}{0}
\section{Gadobutrol - NV centre Interaction}

\begin{figure}
    \includegraphics[width=0.8\linewidth]{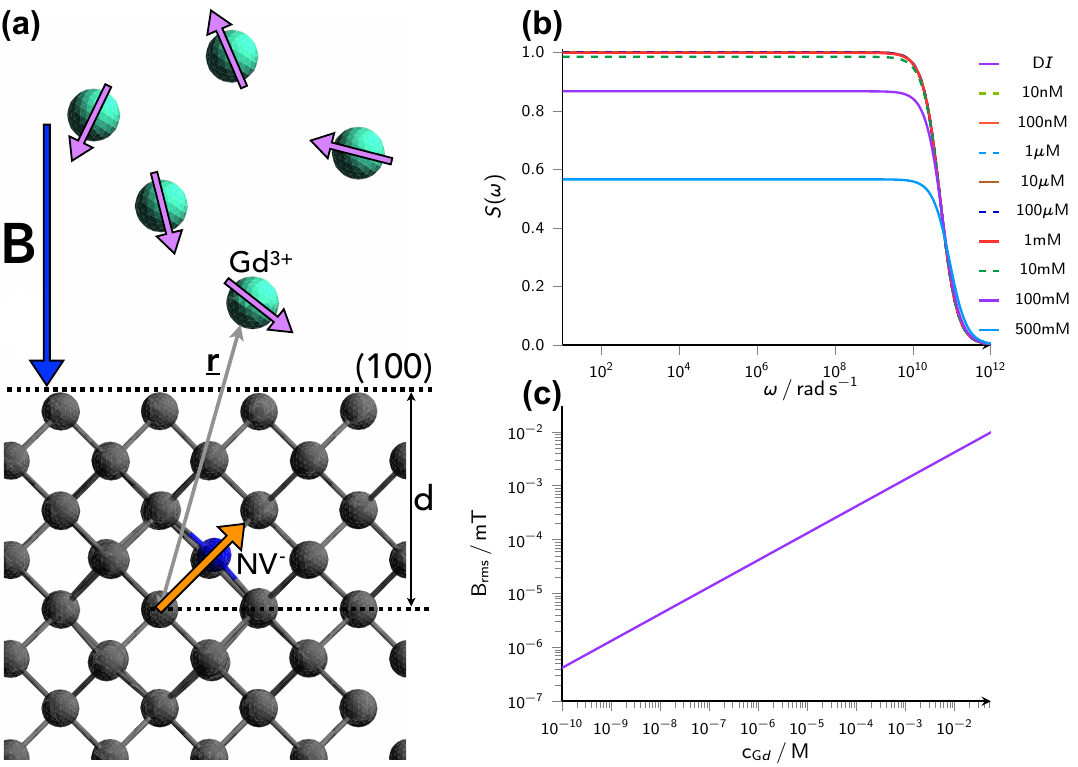}
    \caption{Magnetic noise from Gd$^{3+}$ ions near the diamond surface. (a) A schematic displaying the Gd$^{3+}$ ions relative to the surface of the diamond, cut along the $(100)$ plane with the NV centre at distance $d$ below the surface, and with an applied magnetic field perpendicular to the surface. (b) The magnetic noise spectral density of Gd$^{3+}$ calculated according to Eq.~\eqref{eq:spec_dens} for the concentrations of gadobutrol considered in this work. (c) The root-mean-squared magnetic field strength at an NV centre 5 nm below the diamond surface as a function of gadobutrol concentration, calculated with Eq.~\eqref{eq:b2_gd}. Together, (b) and (c) show that increasing gadobutrol concentration both broadens the magnetic noise spectral density and increases the root-mean-squared field strength experienced by a nearby NV centre. Both effects increase the NV centre's longitudinal relaxation rate, $\Gamma_{1}(\mathbf{B})$, Eq.~\eqref{eq:long_rlx_rate}, reducing the achievable spin polarisation and thereby decreasing the ODMR and MQ signal with increasing gadobutrol concentration, as observed in the main text.}
    \label{fig:NV_GD_Schematic}
\end{figure}

Gd$^{3+}$ ions in gadobutrol solution generate fluctuating magnetic fields even in the absence of an external magnetic field. These fluctuations originate from the ions’ magnetic dipole moments, which arise due to their 4f$^7$ valence electron configuration and a spin quantum number of $S = \tfrac{7}{2}$. The fluctuating field arises due to the magnetic dipolar interaction between Gd$^{3+}$ ions, translational diffusion, rotation and vibration within their chemical environment.~\cite{Du2024} The interaction between an NV centre spin, $\mathbf{S}_{i}$ , located at a depth $d$ below the diamond surface and a Gd$^{3+}$ spin, $\mathbf{S}_{j}$, in solution on the surface, separated by a distance  $\mathbf{r}_{ij}$, is described by the interaction Hamiltonian:

\begin{equation}\label{eq:dipole_dipole}
    \hat{H}_{ij} = \frac{\mu_{0}\hbar}{4\pi}\left(\frac{g_{i}\mu_{B}}{\hbar}\right)\left(\frac{g_{j}\mu_{B}}{\hbar}\right)\left\lbrack \frac{\mathbf{S}_{i}\cdot\mathbf{S}_{j}}{\mathbf{r}_{ij}^{3}} - \frac{3(\mathbf{S}_{i}\cdot\mathbf{r}_{ij})(\mathbf{S}_{j}\cdot\mathbf{r}_{ij})}{\mathbf{r}_{ij}^{5}}\right\rbrack
\end{equation}

where $\mu_{0}$ is the permeability of free space, $\mu_{B}$ the Bohr magneton and $g_{i}$ the electronic g-factor of spin $i$. Since the strength of this interaction decays rapidly with separation, this coupling may be approximated through average pairwise interactions, given for a homogeneous spin solution by~\cite{McGuinness2013,Wilson2023}

\begin{equation}\label{SI:eq:hdip_approx}
    \hat{H} \approx \frac{\mu_{0}\mu_{B}^{2} g^{2}_\text{Gd}}{4\pi \hbar}\frac{\mathbf{S}^{2}_\text{Gd}}{\bar{r}^{3}},
\end{equation}

where the average spin separation $\bar{r}$ is proportional to the number of spins $N$ per unit volume $V$, or concentration $c$, as

\begin{equation}
    \bar{r} = \frac{\Gamma(4/3)}{(4\pi/3)^{-1/3}} \left(\frac{N}{V}\right)^{-\frac{1}{3}}.
\end{equation}

The resulting dependence on Gd$^{3+}$ concentration, $c_\text{Gd}$ has been shown to result in a dipole interaction contribution to the spin fluctuation approximated as $f_\text{dip} = \text{c}_\text{Gd}\cdot 77\text{GHz\,M}^{-1}$. \cite{Steinert2013} The diffusional, rotational and vibrational contributions have been shown by Steinert~\textit{et al.} to be well approximated by $f_\text{dif}\sim 40\text{MHz}$, $f_\text{rot}\sim 100\text{MHz}$ and $f_\text{vib}\sim 50\text{GHz}$ respectively. The total fluctuation rate is thus given by~\cite{Steinert2013,Ziem2013,Du2024}

\begin{equation}\label{eq:fluc_rate}
    f_\text{Gd} = f_\text{dip} + f_\text{dif} + f_\text{rot} + f_\text{vib} \sim \text{c}_\text{Gd}\cdot 77\text{GHz\,M}^{-1} + 51.4\text{GHz}.
\end{equation}

Given the random orientations of the Gd$^{3+}$ in the absence of an external field, the magnetic field they induce may be considered spatially isotropic. The variance in transverse magnetic field at an NV centre situated at depth $d$ below the diamond-substrate interface was approximated by Steinert~\textit{et al.} as~\cite{Steinert2013}

\begin{equation}\label{eq:b2_gd}
    \langle B_\text{Gd} \rangle^{2} \sim 2.1 \!\times \! 10^{4}\frac{\pi N_{A} \text{c}_\text{Gd}}{16d^{3}}\left( \frac{\mu_{0}\hbar}{4\pi}\gamma_\text{Gd}\right)^{2},
\end{equation}

where $N_{A}$ is Avogadro's number and $\gamma_\text{Gd}$ is the gyromagnetic ratio of the Gd$^{3+}$ ion.\cite{Steinert2013,Hall2016,Simpson2017} The effective root-mean-squared field strength, $B_{rms}=\langle B_\text{Gd} \rangle^{2}/\gamma_\text{Gd}$, at an NV centre situated 5 nm below the interface is modeled using Eq.~\eqref{eq:b2_gd} in Figure~\ref{fig:NV_GD_Schematic}(c). This demonstrates the increase in the magnitude of magnetic fluctuations with increasing gadobutrol concentration discussed in the main text. The magnetic fluctuation spectral density function at the NV centre is given by

\begin{equation}\label{eq:spec_dens}
    S(\omega) = \frac{\gamma_\text{NV}^{2}\langle B_\text{Gd} \rangle^{2}}{\pi} \frac{f_\text{Gd}}{f_\text{Gd}^{2}+(\omega-\omega_\text{Gd})^{2}},
\end{equation}

where $\omega_\text{Gd}=\gamma_\text{Gd}B$ is the Larmor frequency of the Gd$^{3+}$ ion at field strength $B$. Figure~\ref{fig:NV_GD_Schematic}(b) shows the spectral function's dependence on ion concentration. The concentration-dependent broadening of the spectral density indicates that Gd$^{3+}$ ions introduce a wider range of magnetic field frequencies, leading to more efficient spin relaxation in NV centres. Finally, the longitudinal relaxation rate $\Gamma_{1}$ of the NV centre in the presence of an external magnetic field ($\mathbf{B}$) and paramagnetic spin noise is calculated by integrating the spectral density function with an NV filter function,

\begin{equation}\label{eq:long_rlx_rate}
    \Gamma_{1}(\mathbf{B}) = \Gamma_{1,\text{int}} + \int \frac{\Gamma_{2}}{2(\Gamma_{2}^{2} + (D_{gs} \pm \gamma_{NV}B_{\parallel} - \omega)^{2}}S(\omega)\mathrm{d}\omega,
\end{equation}

where $\Gamma_{1,\text{int}}$ is the intrinsic relaxation rate of the NV centre, $\Gamma_{2}$ is its dephasing rate, $B_{\parallel}$ is the component of the external field parallel to the NV centre quantization axis and $D_{gs}$ is the axial zero-field splitting parameter.\cite{Simpson2017,Du2024} As the NV centre relaxes, the population of the lowest energy state decays with time $t$ as $P=\frac{1}{6}(2+\exp(-\Gamma_{1}^{-}t)+\exp(-\Gamma_{1}^{+}t) + 2\exp(-(\Gamma_{1}^{-}+\Gamma_{1}^{+})t))$. An increase in Gd$^{3+}$ ion concentration strengthens magnetic fluctuations, leading to an increased longitudinal relaxation rate. This increased relaxation rate causes a more rapid decay in the population of the lowest energy component of the ground state towards equilibrium.

\clearpage
\section{Nanotube Analysis}
In addition to MQ and ODMR analysis presented in the main text, all nanotubes were subject to further characterisation by Transmission Electron Microscopy (TEM), Energy Dispersive X-ray spectroscopy (EDX, Cu is from the TEM grid), Thermogravimetric Analysis (TGA), Powder X-ray Diffraction (PXRD), and Electron Paramagnetic Resonance (EPR) spectroscopy. TGA was used to determine the amount of catalyst in the SWCNT samples, with residual weight analysis confirming that SWCNT1 has a lower percentage metal loading than SWCNT2 (5.1 \% in Figure S3f vs 16.4 \% in Figure S4f). BNNTs possess very little paramagnetic character, as evidenced by their near silent EPR spectrum compared to both SWCNT samples. The BNNTs in this study contained no residual metal catalyst, as confirmed by PXRD (Figure S2e) and EDX (Figure S2c).

\begin{figure}
    \centering
    \includegraphics[width=\linewidth]{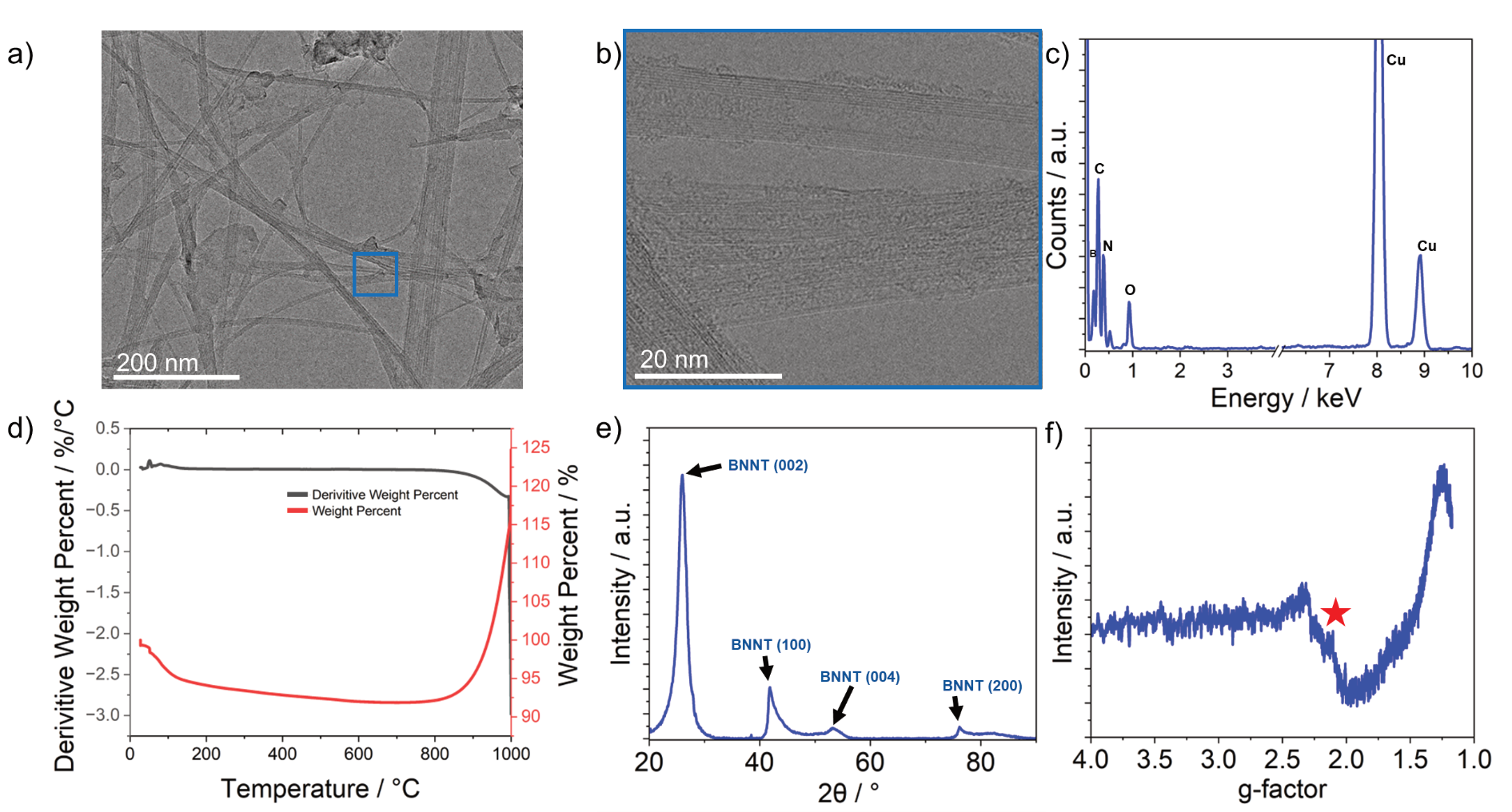}
    \caption{Nanoscale and Bulk analysis of BNNTs. a) 200 kV TEM image of bundles of BNNTs, b) magnified 200 kV TEM image of the BNNTs shown in the blue area in a), c) EDX analysis of the BNNTs shown in a), d) TGA thermogram of BNNTs heated in air, e) PXRD analysis of BNNTs, with the characteristic reflections indexed, f) EPR analysis of BNNTs. The red star shows absorption corresponding to the small concentration of paramagnetic defects found in pristine BNNTs, previously reported by Panich \textit{et al.}\cite{Panich2005} Collectively, the analysis in this figure confirms that the BNNT sample exhibits only weak paramagnetic character, consistent with the relatively small spin-noise response observed in the ODMR and MQ measurements.}  
    \label{BNNT}
\end{figure}

\begin{figure}
    \centering
    \includegraphics[width=\linewidth]{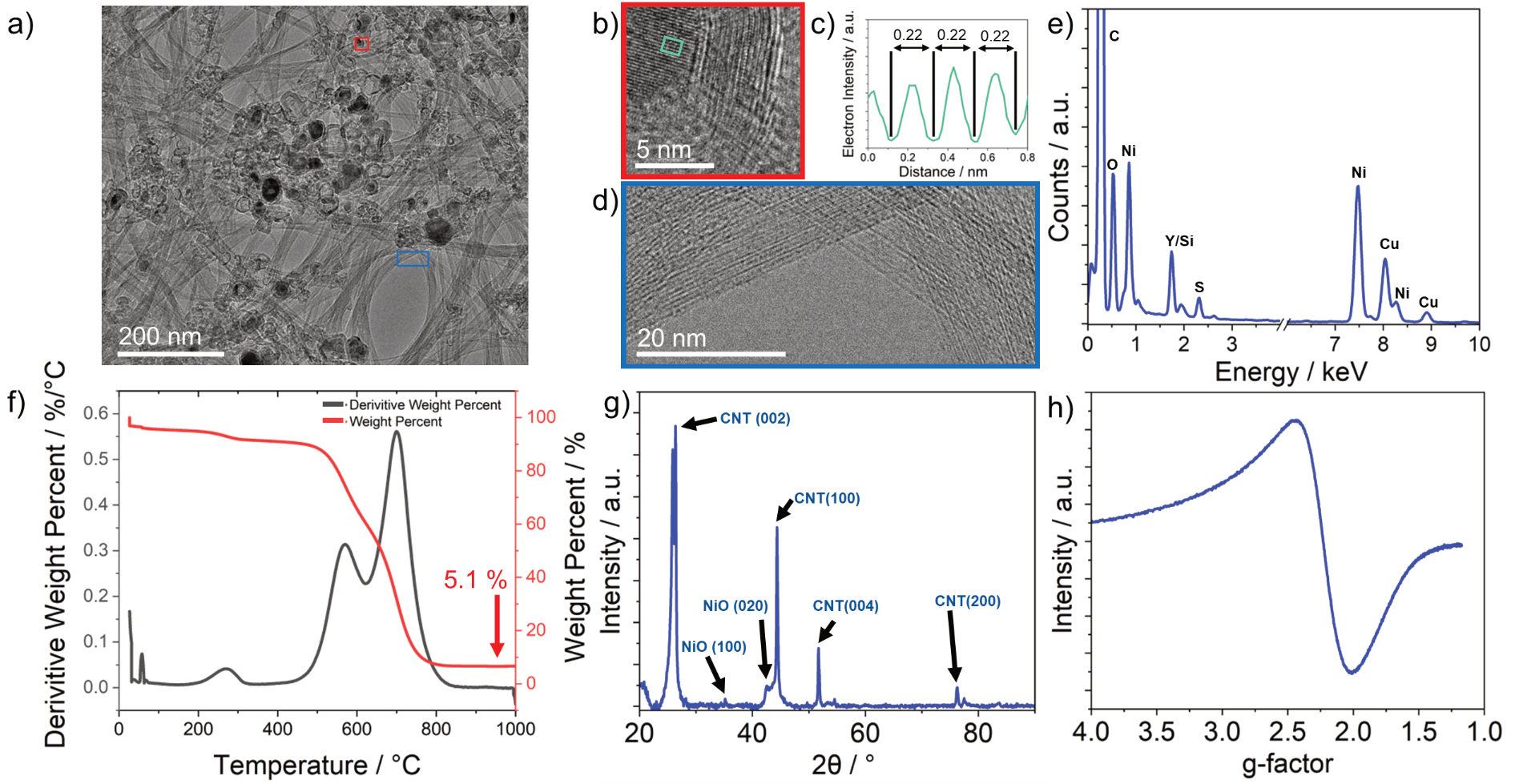}
    \caption{Nanoscale and Bulk analysis of SWCNT 1. a) 200 kV TEM image of bundles of SWCNT1, b) magnified 200 kV TEM image of the area shown in red in a), c) line profile analysis of the area highlighted in cyan in a), showing spacings corresponding to the (002) plane of NiO, d) magnified 200 kV TEM image of the area shown in blue in a), e) EDX analysis of the SWCNTs shown in a) confirming the presence of the residual Ni-based catalyst from industrial synthesis, f) TGA thermogram of SWCNT1 heated in air showing a remaining 5.1 \% weight percent due to the residual Ni catalyst, g) PXRD analysis of SWCNT1, h) EPR analysis of SWCNT1, note that the EPR spectrum demonstrates substantially greater spin-active character than observed for the BNNT sample. Collectively, the analysis in this figure confirms SWCNT1 contains residual Ni catalyst ($\sim$5.1 \% weight percent), resulting in significantly greater paramagnetic character than BNNTs and providing a likely source of the enhanced spin-noise response observed in the ODMR and MQ measurements.}
    \label{SWCNT1}
\end{figure}
\begin{figure}
    \centering
    \includegraphics[width=\linewidth]{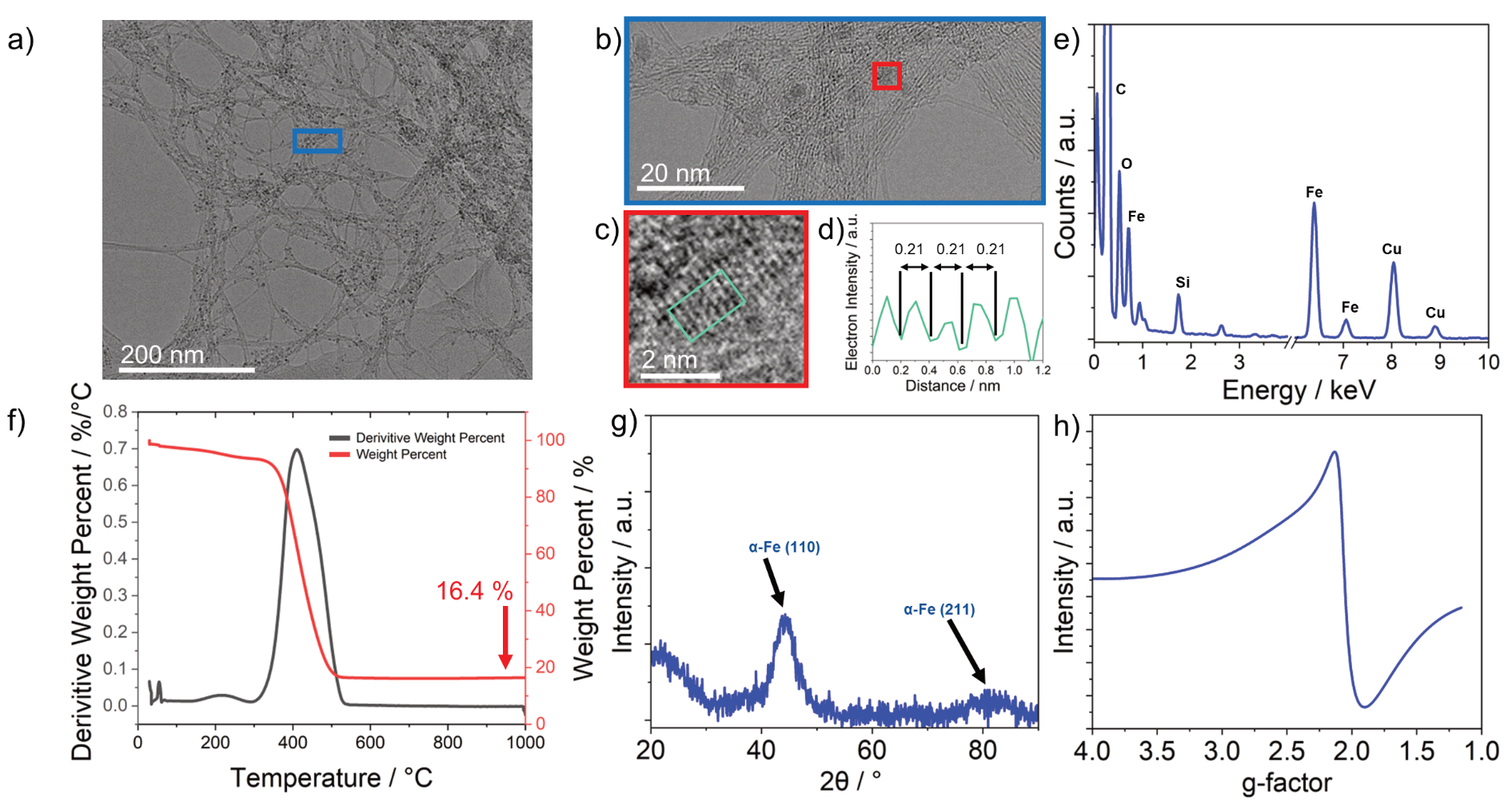}
    \caption{Nanoscale and Bulk analysis of SWCNT2 a) 200 kV TEM image of bundles of SWCNT2, b) magnified 200 kV TEM image of the area shown in blue in a), c) magnified 200 kV TEM image of the area shown in red in b), d) line profile analysis of the cyan area shown in c), showing spacings corresponding to the (110) plane of $\alpha$-Fe, e) EDX analysis of the SWCNTs shown in a) showing the presence of residual Fe catalyst from industrial synthesis, f) TGA thermogram of SWCNT2 heated in air showing a remaining 16.4 \% weight percent due to the residual Fe catalyst, g) PXRD analysis of SWCNT2 showing reflections assigned to $\alpha$-Fe. No reflections corresponding to SWCNTs were observed, which is attributed to the lack of bundling/graphitic stacking in these samples. h) EPR analysis of SWCNT2, demonstrating the strong paramagnetic response of SWCNT2. Collectively, the analysis in this figure confirms that SWCNT2 contains a high residual Fe catalyst loading ($\sim$16.4 \%) and exhibits the strongest paramagnetic character of the nanotube samples studied, providing a likely source of the enhanced spin-noise response observed in the ODMR and MQ measurements.}
    \label{SWCNT2}
\end{figure}

\clearpage
\section{Diamond Cleaning}
All diamond plates were cleaned prior to any measurements presented in the main text following the same procedure. The diamond plates were first treated with a 1:1 ratio of concentrated sulfuric and nitric acid for 2 hours in a pressure vial. The acid mixture was then neutralised with sodium hydroxide, rinsed with MilliQ water, and air oxidised at 465 \textdegree C for 2 hours in a Thermo Thermolyne benchtop oven with a temperature ramp of 3 \textdegree C $/ $min to 6 \textdegree C $/ $min.  The plates were then refluxed in acid for 2 hours, neutralised, boiled in MilliQ water for 1 hour, and stored in MilliQ water until use. 

\section{Instrumentation}
\subsection{Optics}
\begin{figure}
\centering
    \includegraphics[width=\linewidth]{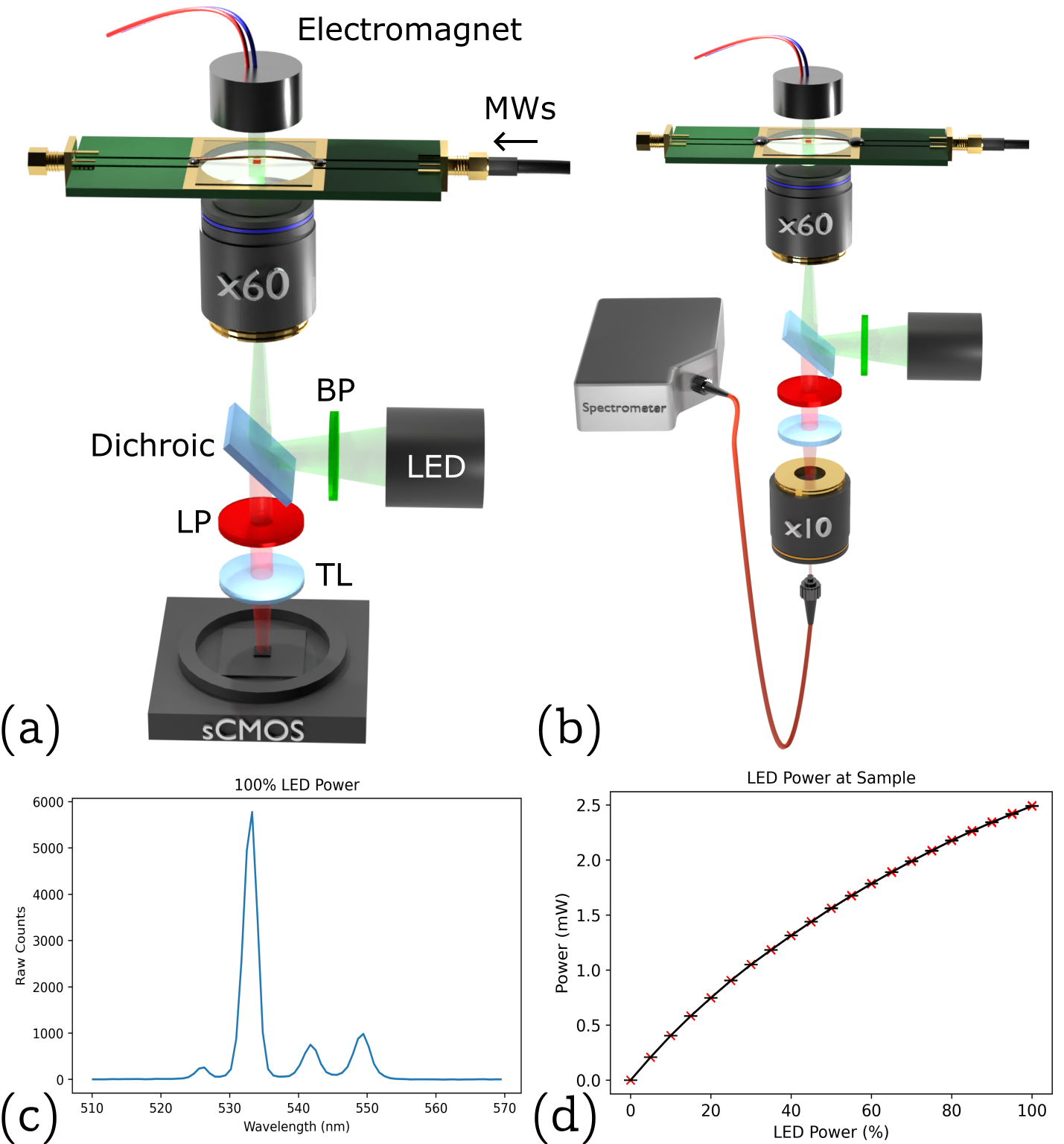}
    \caption{Optical setup and characterisation of the excitation source.(a) Schematic of the inverted fluorescence microscope's key optical components for NV sensing. BP: bandpass filter, LP: long pass filter, TL: tube lens, MW: microwaves (for ODMR), (b) Spectrophotometry setup showing modified optical components from (a), (c) Spectral profile of the CoolLED-pE4000 green excitation light at the sample (following 530–550 nm bandpass filtering). (d) Optical power at the sample plane vs. LED drive current setting (measured with an Excelitas X-cite power measurement system)}
    \label{optics}
\end{figure}

All ODMR and MQ measurements were performed with an inverted microscope with a green LED as the illumination source. The key optical components in the excitation and detection paths are shown in Figure \ref{optics}(a). The spectral profile of the CoolLED-pE4000 excitation source operated in '550 nm' mode following 530 - 550 nm bandpass filtering and direction towards the sample plane with a 570 nm dichroic filter is shown in Figure \ref{optics}(b). Finally, optical power measurements of the excitation power at the sample plane are presented in Figure \ref{optics}(d).

\subsection{Electromagnet}
\begin{figure}
\centering
    \includegraphics[width=\linewidth]{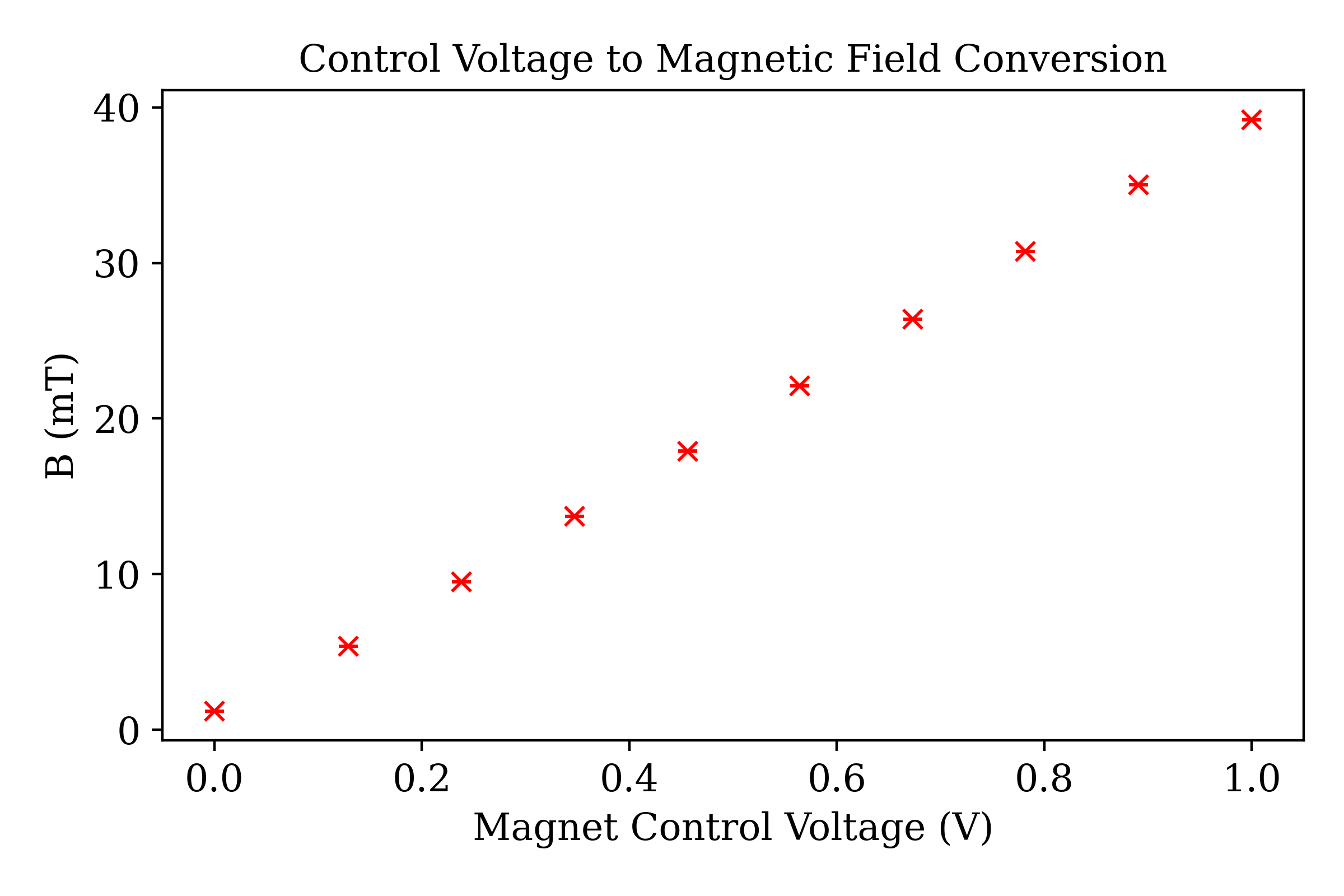}
    \caption{Characterisation of the electromagnet magnetic field. Magnetic field strength as measured by a transverse Hall probe, at control voltages used with 3 mm spacing between the electromagnet face and the Hall probe.}
    \label{Bvals}
\end{figure}

A transverse Hall probe (Hirst GM08 handheld Gaussmeter equipped with TP002 Probe) was used to calibrate the electromagnet - enabling the conversion from the amplitude of the square wave modulation to magnetic flux density generated by the magnet during MQ measurements. The transverse Hall probe was placed in the microscope sample plane, and the linear translation stage was used to position the magnet in the imaging position, 3 mm above the sample. Figure \ref{Bvals} shows the measured field amplitude for a range of magnet control voltages used in the main text. 

\subsection{Spectral MQ Signal acquisition}

\begin{figure}
\centering
    \includegraphics[width=\linewidth]{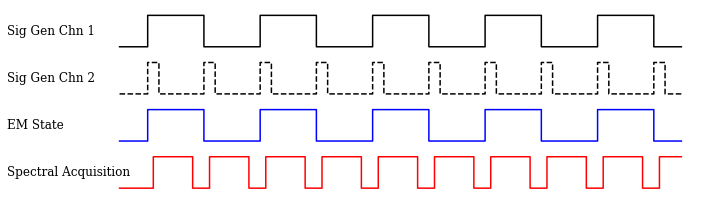}
    \caption{Timing sequence for synchronised spectral MQ measurements. Timing diagram showing the synchronisation of electromagnet modulation and spectrophotometer acquisition using logic pulses. Black traces show 3.3 V signal-generator outputs, the blue trace indicates the electromagnet drive state (high: current supplied; low: no drive), and the red trace indicates spectrophotometer integration (high: integrating; low: inactive).}    \label{spec_time}
\end{figure}

In MQ spectral sensing experiments a single spectrum was acquired per magnetic low and magnetic high state. The spectrometer was preconfigured to acquire a 350 ms spectrum after the rising edge of a TTL trigger signal. The experimental timing was controlled using the Tektronix AFG, where one channel generated the analog signal to modulate the magnetic field of the electromagnet, and the second channel generated the trigger signal for the spectrometer. An acquisition time of 350 ms per spectrum was chosen so that the spectra would be acquired while the magnet was in a single magnetic state. The timing diagram is shown in Figure \ref{spec_time}.

\clearpage
\section{Data Processing}

\subsection{Spatially Uniform Samples}

\subsubsection{Spectrophotometry}
Spectrophotometry was performed on spatially uniform samples (bare plate measurements and Gadobutrol titrations). A spectrum was acquired each time the electromagnet changed state. The 'magnet high' and 'magnet low' spectra presented here are the mean average of all acquired spectra in a measurement. For the difference spectra, the difference (magnet low minus magnet high) was calculated for sequentially acquired on-off pairs, and the mean average of all pairs was then determined and presented as the 'difference spectrum' for a given measurement.

\subsection{Nanotube Spatial Imaging}
For nanotube imaging, ODMR signal was determined pixel-wise from the average of ten repeated ODMR sweeps, followed by 5 x 5 binning (600 x 600 to 120 x 120, for SNR). The ODMR signal was the average amplitude of the fitted Lorentzians. The same pixel-wise measurements were performed on MQ data. For MQ, the average of the magnet-high state frames and the magnet-low state frames was then determined, and the MQ signal at each pixel was computed from subtracting the mean magnet-high value from the mean magnet-low frame value.

\section{Theoretical Methods: Additional Details}

\subsection{Detailed Spin Hamiltonian}

The zero-field splitting (ZFS) tensor is symmetric and traceless, and on rotation into the coordinate frame in which it is diagonal, Eq. (1) of the main text may be expressed more simply as

\begin{equation}
    \hat{\mathcal{H}}=D[\hat{S}_{z}^{2}-\frac{1}{3}S(S+1)]+E(\hat{S}_{x}^{2}-\hat{S}_{y}^{2})+\mu_{B}\mathbf{B}\cdot \mathbf{g}\cdot\hat{S}
\end{equation}

where the ZFS is uniquely defined by the axial ZFS $D=D_{zz}-\frac{1}{2}(D_{xx}+D_{yy})$ and rhombic ZFS $E=\frac{1}{2}(D_{xx}-D_{yy})$ parameters.~\cite{Sinnecker2006,Mostafanejad2014,Chibotaru2014} In the ground state equilibrium geometry of the $NV^{-}$ centre, $E=0$ due to symmetry, hence the $M_{s}=\pm1$ sublevels of the triplet states are mutually degenerate but separated in energy from the $M_{s}=0$ sublevel by D, as shown in Figure 6(a) of the main text.~\cite{Mostafanejad2014,Chibotaru2014}
%[cite: 225, 226, 227] -> ~\cite{Sinnecker2006,Mostafanejad2014,Chibotaru2014}
%[cite: 227, 228] -> ~\cite{Mostafanejad2014,Chibotaru2014}

\subsection{Mathematical Description of Spin Mixing}

For a given multiplet state, the spin sublevels $|SM_{s}\rangle$ form a basis in which the eigenstates of the spin Hamiltonian are represented,

\begin{equation}\label{eq:spinmixingsublevels}
\mathcal{\mathcal{H}}|\tilde{i}\rangle=E_{i}|\tilde{i}\rangle \text{ where } |\tilde{i}\rangle=\sum_{M_{s}=-S}^{S}c_{\tilde{i},M_{s}}|SM_{s}\rangle
\end{equation}

In the absence of a magnetic field, the spin Hamiltonian for the $NV^{-}$ centre reduces to $\hat{\mathcal{H}}=D[\hat{S}_{z}^{2}-\frac{1}{3}S(S+1)]$; it follows that the Hamiltonian matrix $\langle SM_{s}|\mathcal{H}|SM_{s}^{\prime}\rangle$ is diagonal and the spin states $|SM_{s}\rangle$ are eigenstates of the spin Hamiltonian. In the presence of a magnetic field aligned along the zero-field axis of quantization, taken to be the z-axis, the spin Hamiltonian for the NV centre becomes $\hat{\mathcal{H}}=D[\hat{S}_{z}^{2}-\frac{1}{3}S(S+1)]+\mu_{B}B_{z}g_{zz}\hat{S}_{z}$, for which the Hamiltonian matrix remains diagonal. However, in the presence of a magnetic field not aligned to the axis of quantization, the Hamiltonian in the basis of the spin sublevels is no longer diagonal and its eigenstates become linear combinations of the zero-field spin states, $|\tilde{i}\rangle$, as shown in Figure 6(b) of the main text and given in Eq.~\eqref{eq:spinmixingsublevels}.~\cite{Chibotaru2014,Segawa2023}
%[cite: 228, 229, 230, 231, 232, 233] -> ~\cite{Chibotaru2014,Segawa2023}

\subsection{Additional Energy Level Details}

Figure 6(a) of the main text shows the transitions between electronic energy levels at zero field; optical excitations from ${}^{3}A_{2}$ to ${}^{3}E$, broadband photoluminescence upon de-excitations from ${}^{3}E$ back to ${}^{3}A_{2}$, along with non-radiative transitions from ${}^{3}E$ to ${}^{1}A_{1}$ and ${}^{1}E$ to ${}^{3}A_{2}$.~\cite{Choi2012} The optical excitation and radiative decay processes are spin conserving whilst the non-radiative decay processes between singlet and triplet states are spin non-conserving; these inter-system crossing (ISC) transitions, enabled by spin-orbit coupling and electron-phonon interactions,~\cite{Thiering2017,Razinkovas2021} are highly state selective and occur from ${}^{3}E^{\pm1}\rightarrow{}^{1}A_{1}$ at a far higher rate than from ${}^{3}E^{0}\rightarrow{}^{1}A_{1}$.~\cite{Goldman2015} The transition rates between states are generally approximated by Fermi's Golden Rule,

\begin{equation}\label{eq:TransitionRatesFermi}
%    \Gamma_{\tilde{i}\rightarrow\tilde{j}} = \frac{2\pi}{\hbar} \sum_{a} |\langle \tilde{j} | \hat{\mathcal{H}}^{\prime} | \tilde{i} \rangle |^{2} P_{a}(T) (1 - P_{a}(T)) \delta(E_{\tilde{i}} - E_{\tilde{j}} \pm \hbar\omega),
    \Gamma_{\tilde{i}\rightarrow\tilde{j}} = \frac{2\pi}{\hbar} \vert\langle \tilde{j} \vert \hat{\mathcal{H}}'\vert \tilde{i} \rangle \vert^{2} \sum_{b} P_{b}(T)\vert \langle \nu^{\tilde{j}}_{b} \vert \nu_{0}^{\tilde{i}}\rangle \vert^{2} \delta (E_{\tilde{j}}+\epsilon_{b}^{\tilde{j}}-E_{\tilde{i}}-\epsilon_{a}^{\tilde{i}}\pm\hbar\omega),
\end{equation}

where $\nu_{a}^{\tilde{i}}$ is the $a^{th}$ phonon mode in electronic state $|\tilde{i}\rangle$, with $\epsilon_{a}^{\tilde{i}}$ and $E_{\tilde{i}}$ respectively, with population at temperature T given by $P_{a}(T)$, $\hat{\mathcal{H}}^{\prime}$ is the perturbing Hamiltonian of the transition; the electronic dipole operator $\hat{\mu}$ for radiative transitions absorbing or emitting a photon of angular frequency $\omega$ and the spin-orbit coupling operator $\hat{\mathcal{H}}_{SO}$ for non-radiative ISCs $(\omega=0)$.

At zero field, the non-radiative transition rate from the $M_{s}=\pm1$ sublevels of ${}^{3}E$ to the ${}^{1}A_{1}$ singlet state is significantly higher than from the $M_{s}=0$ state, however, the non-radiative transition rate from the ${}^{1}E$ singlet state to the ${}^{3}A_{2}$ ground state is somewhat higher for the $M_{s}=0$ sublevel than the $M_{s}=\pm1$ sublevels. As a result, repeated optical excitation with an off-resonant laser light results in a high degree of spin polarisation of the ground state into the $M_{s}=0$ spin sublevel.~\cite{Choi2012,Goldman2015} In the presence of an off-axis magnetic field, the mixing of spin sublevels as given in Eq.~\eqref{eq:spinmixingsublevels} leads to changes in the relative rates of radiative and non-radiative transition which may be approximated as~\cite{Tetienne2012}
%[cite: 240, 241, 242, 243, 244, 245, 246] -> ~\cite{Choi2012,Goldman2015,Tetienne2012}

\begin{equation}\label{eq:RelativeRates}
    \Gamma_{\tilde{i}\rightarrow\tilde{j}} = \sum_{M_{s}=-S}^{S} \sum_{M_{s}'=-S'}^{S'} c_{\tilde{i},M_{s}} c_{\tilde{j},M_{s}'}^{\ast} \Gamma_{M_{s} \rightarrow M_{s}'}
\end{equation}

As such, the population accumulation into the lowest-energy component of the ground state is much less efficient under a magnetic field, with a higher rate of non-radiative decay from the lowest-energy components of the excited triplet state relative to zero field, thus decreasing the photoluminescence intensity.

\bibliography{SI_ref}